\documentclass{vldb}
\usepackage{balance}  
\usepackage{graphicx} 
\usepackage{url}      
\usepackage{amsmath}
\usepackage{color}
\usepackage{cancel}
\usepackage{listings}
\usepackage[normalem]{ulem}
\usepackage{graphicx}
\usepackage{subcaption}
\usepackage{booktabs}
\usepackage{graphicx}
\usepackage[ruled,vlined,algonl,boxed]{algorithm2e}
\usepackage{algorithmic}
\usepackage{wrapfig}
\usepackage{enumitem}
\usepackage{xspace}
\usepackage{xcolor}
\usepackage{colortbl}
\usepackage[T1]{fontenc}

\usepackage[nocompress]{cite}
\usepackage{microtype}
\usepackage[section]{placeins}

\makeatletter
\def\url@leostyle{
  \@ifundefined{selectfont}{\def\UrlFont{\sf}}{\def\UrlFont{\small\bf\ttfamily}}}
\makeatother
\makeatletter
\def\@copyrightspace{\relax}
\makeatother

\def\pprw{8.5in}
\def\pprh{11in}

\newtheorem{definition}{Definition}

\newtheorem{example}[definition]{Example}

\usepackage[pdftex]{hyperref}
\hypersetup{
  colorlinks=false,
  linkcolor=darkred,
  citecolor=darkgreen,
  urlcolor=darkblue
}

\definecolor{light-gray}{gray}{0.95}
\definecolor{mid-gray}{gray}{0.85}
\definecolor{darkred}{rgb}{0.7,0.25,0.25}
\definecolor{darkgreen}{rgb}{0.15,0.55,0.15}
\definecolor{darkblue}{rgb}{0.1,0.1,0.5}
\definecolor{blue}{rgb}{0.19,0.58,1}

\newcommand{\red}[1]{\textcolor{red}{#1}}


\newcommand{\techreport}[1]{#1}

\newcommand{\stitle}[1]{\vspace{0.5em}\noindent\textbf{#1}}

\makeatletter
\def\maketag@@@#1{\hbox{\m@th\normalfont\normalsize#1}}
\DeclareRobustCommand*\textsubscript[1]{
          \@textsubscript{\selectfont#1}}
        \def\@textsubscript#1{
          {\m@th\ensuremath{_{\mbox{\fontsize\sf@size\z@#1}}}}}
\makeatother

\newcommand{\sys}{QFix\xspace}
\newcommand{\naive}{\emph{basic}\xspace}

\newcommand{\incremental}{\sys-{\it inc}\xspace}
\newcommand{\dt}{DecTree\xspace}

\begin{document}

\title{{\sys}: Diagnosing errors through query histories}

\numberofauthors{3}
 \author{
  \alignauthor Xiaolan Wang\\
    \affaddr{School of Computer Science}\\
    \affaddr{University of Massachusetts}\\
    \email{xlwang@cs.umass.edu}\\
  \alignauthor Alexandra Meliou\\
  \affaddr{School of Computer Science}\\
    \affaddr{University of Massachusetts}\\
    \email{ameli@cs.umass.edu}\\
  \alignauthor Eugene Wu\\
    \affaddr{Computer Science}\\
    \affaddr{Columbia University}\\
    \email{ewu@cs.columbia.edu}\\
}

\maketitle

\begin{abstract}
    \looseness -1
Data-driven applications rely on the correctness of their data to
function properly and effectively. Errors in data can be incredibly
costly and disruptive, leading to loss of revenue, incorrect
conclusions, and misguided policy decisions. While data cleaning tools
can purge datasets of many errors before the data is used,
applications and users interacting with the data can introduce new
errors. Subsequent valid updates can obscure these errors and
propagate them through the dataset causing more discrepancies. Even
when some of these discrepancies are discovered, they are often
corrected superficially, on a case-by-case basis, further obscuring
the true underlying cause, and making detection of the remaining
errors harder.

In this paper, we propose \sys, a framework that derives explanations
and repairs for discrepancies in relational data, by analyzing the
effect of queries that operated on the data and identifying potential
mistakes in those queries. \sys is flexible, handling scenarios where
only a subset of the true discrepancies is known, and robust to
different types of update workloads. We make four important
contributions: (a) we formalize the problem of diagnosing the causes
of data errors based on the queries that operated on and introduced
errors to a dataset; (b)
we develop exact methods for deriving diagnoses and fixes for
identified errors using state-of-the-art tools; (c) we present several
optimization techniques that improve our basic approach without
compromising accuracy, and (d) we leverage a tradeoff between accuracy
and performance to scale diagnosis to large datasets and query logs,
while achieving near-optimal results. We demonstrate the effectiveness
of \sys through extensive evaluation over benchmark and synthetic
data.

\end{abstract}

\section{Introduction}
\label{s:intro}

Poor data quality is a hard and persistent problem.  
It is estimated to cost the US economy more than \$600 billion
per year~\cite{eckerson2002} and erroneous price data in retail databases
alone cost the US consumers \$2.5 billion each year~\cite{Fan2008}. 
Although data
cleaning tools can purge many errors from a dataset before downstream 
applications use the data, datasets can frequently change as applications
and users execute queries that modify the data.
Mistakes in these queries can introduce errors to the data, and these
errors can propagate to more data by subsequent update queries.
By the time errors are detected, their origin has often been obscured and it is difficult to identify the offending query and correct it.
Identifying and correcting errors in the data directly is suboptimal, as it targets the symptom,
rather than the underlying cause. Fixing the manifested data errors on a
case-by-case basis often obscures the root of the problem and other data that may have been
affected. Therefore, traditional data cleaning approaches are not well-suited
for this setting: While they provide general-purpose tools to identify and
rectify anomalies in the data, they are not designed to diagnose the causes of
errors that are rooted in erroneous updates.
Some data cleaning systems try to identify structural sources of
mistakes~\cite{wang2015}, but they are unable to trace the source of
the mistakes to particular faulty queries.

While improving data quality and correcting data errors has been an important
focus for data management research, handling new errors, introduced during
regular database interactions, has received little attention. Most work in
this direction focuses on \emph{guarding against} erroneous updates. For
example, integrity constraints~\cite{Khoussainova2006} reject some improper
updates, 
but only if the data falls outside rigid, predefined ranges.
Certificate-based verification~\cite{Chen2011} is less rigid, but it is
impractical and non-scalable as it requires users to answer challenge
questions before allowing the updates, and it is not applicable to updates
initiated by applications.

\looseness -1
In this paper, we present \sys, a \emph{diagnosis and repair} framework for data errors
caused by erroneous updates. In contrast to existing approaches in data
cleaning that aim to detect and correct errors in the data directly, the goal
of \sys is to identify errors in the queries that introduced errors in the
data. These diagnoses \emph{explain} how errors were introduced to a
dataset, and allow an administrator to easily identify and further validate the most likely query-based sources of these errors. 
In addition, once the erroneous queries have been confirmed, repairing the source of the errors
can potentially lead to the identification of additional discrepancies in
the data that would have otherwise remained undetected.  
We describe two motivating examples; the first is a real-life scenario, provided to us by a large US-based wireless provider.
\begin{example}[Wireless discount policies]\label{ex:telco}

A wireless provider offers company discounts as incentives for
corporate customers. There are different types of discounts (flat, percentage,
fee-based), and their details are specific to corporate agreements. The large
number of policies and complexities in their rules frequently cause policies
to be set incorrectly, leading to errors in the application of discounts to
customers' accounts.

\looseness -1
Customers who notice billing errors contact the provider, but the call centers
do not have the capacity or ability to investigate the complaints deeply. The
standard course of action is to correct mistakes on a case-by-case
basis for each complaint. As a result, unreported errors remain in the
database for a long time, or they never get corrected, and their cause becomes
harder to trace as further queries modify the database.

\end{example}
\begin{example}[Tax bracket adjustment]\label{ex:taxes}
    
Tax brackets determine tax rates for different income levels and are
often adjusted. Accounting firms implement these changes to their
databases by appropriately updating the tax rates of their customers. Mistakes
in these update queries (e.g., Figure~\ref{fig:example}) result in errors in
the tax rates and computed taxes. 

\end{example}
In these application scenarios, data errors are typically reported to
a customer service department, which does not have the resources nor
the capability to investigate the errors more broadly. Instead, errors
are resolved on a case-by-case basis. The goal of \sys is to identify
the query or queries that caused the errors and propose corrections to
those queries.  Once these repairs have been validated (say, by an expert), they can be used to identify unreported
errors and to prevent the introduction of more errors. This problem
has the following important characteristics that render it very difficult, and unsuitable for traditional
techniques:

\begin{description}[leftmargin=*, topsep=0mm, itemsep=0mm]
    
    \item[Obscurity.] \looseness -1
    Handling data errors directly often
    leads to partial fixes that further complicate the eventual diagnosis and
    resolution of the problem. For example, a transaction implementing a
    change in the state tax law updated tax rates using the wrong rate,
    affecting a large number of consumers. This causes a large number of
    complaints to a call center, but each customer agent usually fixes each
    problem individually, which ends up obscuring the source of the problem.
    
    \item[Large impact.] Erroneous queries cause errors at a large scale. The
    potential impact of the errors is high, as manifested in several
    real-world cases~\cite{Yates10, Grady13, sakalerrors}. Further, errors
    that remain undetected for a significant amount of time can instigate
    additional errors, even through valid updates. This increases both their
    impact, and their obscurity.
    
    \item[Systemic errors.] The errors created by bad queries are
    \emph{systemic}: they have common characteristics, as they share the same
    cause. The link between the resulting data errors is the query that
    created them; cleaning techniques should leverage this connection to
    diagnose and fix the problem. Diagnosing the cause of the errors, will
    achieve systematic fixes that will correct all relevant errors, even if
    they have not been explicitly identified.
    
\end{description}
\sys does not replace traditional data cleaning methods, but rather, complements them.
Instead of identifying errors in the data directly, 
\sys targets the diagnosis, explanation, and repair of errors at the root by
leveraging example errors acquired from 
users, traditional data cleaning, or detection techniques.

Diagnosing data errors stemming from incorrect updates is fundamentally
challenging: the search space of possible mistakes and fixes is large, and the
amount of information (number of known errors) may be limited. 
\sys addresses these challenges by analyzing the queries that operated on a
dataset in an efficient and scalable manner. More concretely, we make the
following contributions:

\begin{itemize}[leftmargin=*, topsep=0mm, itemsep=0mm]      
    \item We formalize the problem of diagnosing a set of errors using log
    histories of updates that operated on the data. Given a set of 
    \emph{complaints} as representations of data discrepancies in the current
    state of a dataset, \sys determines how to resolve all of the complaints with the minimal amount of changes to the queries in the log (Section~\ref{sec:abstractions}).
      
    \item We provide an exact error-diagnosis solution through a non-trivial
    transformation of the problem to a mixed integer linear program (MILP) that
    encodes the data provenance of the erroneous tuples. Our approach employs state-of-the-art MILP solvers to identify
    optimal diagnoses that are guaranteed to resolve all complaints without introducing new errors to the data
    (Section~\ref{sec:sol}).
    
    \item We present several optimizations to our basic diagnostic
    method, which reduce the problem size without affecting the
    quality of the produced solutions. Further, we propose an
    incremental repair method that targets the cases where the log
    contains a single corrupted query (or the search focuses on a
    single repair). This incremental analysis of the log allows us to
    scale to large datasets ($100k$ records) and large query logs (hundreds to thousands of update queries). Further, we
    show that our optimization techniques have the additional
    advantage of tolerating incomplete information, such as unreported
    errors (Section~\ref{sec:opt}).

    \item We perform a thorough evaluation of the trade-offs between speed and accuracy of our baseline and optimized methods under a controlled, synthetic setting. In particular, we demonstrate that the  \sys optimizations achieve significant speedup compared to the baseline algorithm (40$\times$ in some of our experimental settings).
    We also evaluate \sys on common OLTP benchmarks and show how \sys can propose fully accurate repairs within milliseconds on a TPC-C workload with $1500$ queries (Section~\ref{sec:experiments}).
\end{itemize}

To the best of our knowledge, \sys is the first system that diagnoses
and repairs errors through query histories. We show that it is
extremely effective and efficient with the update workloads found in
most common benchmarks. 
While \sys trusts its input to be correct, it
can handle incomplete information, and it can be resilient to some
inaccuracies in the reported data errors (Section~\ref{sec:noise}).
\sys does not handle some complex query types
that are less common in update workloads, such as nested queries,
joins, and aggregation. 
It also does not currently deal with large amounts of incorrect
information, such as fake data errors reported by malicious users.
These challenges present exciting future extensions to the system presented in this work.

\section{{\sys} System Architecture}

\begin{figure}[t]
    \centering
        \includegraphics[scale=0.35]{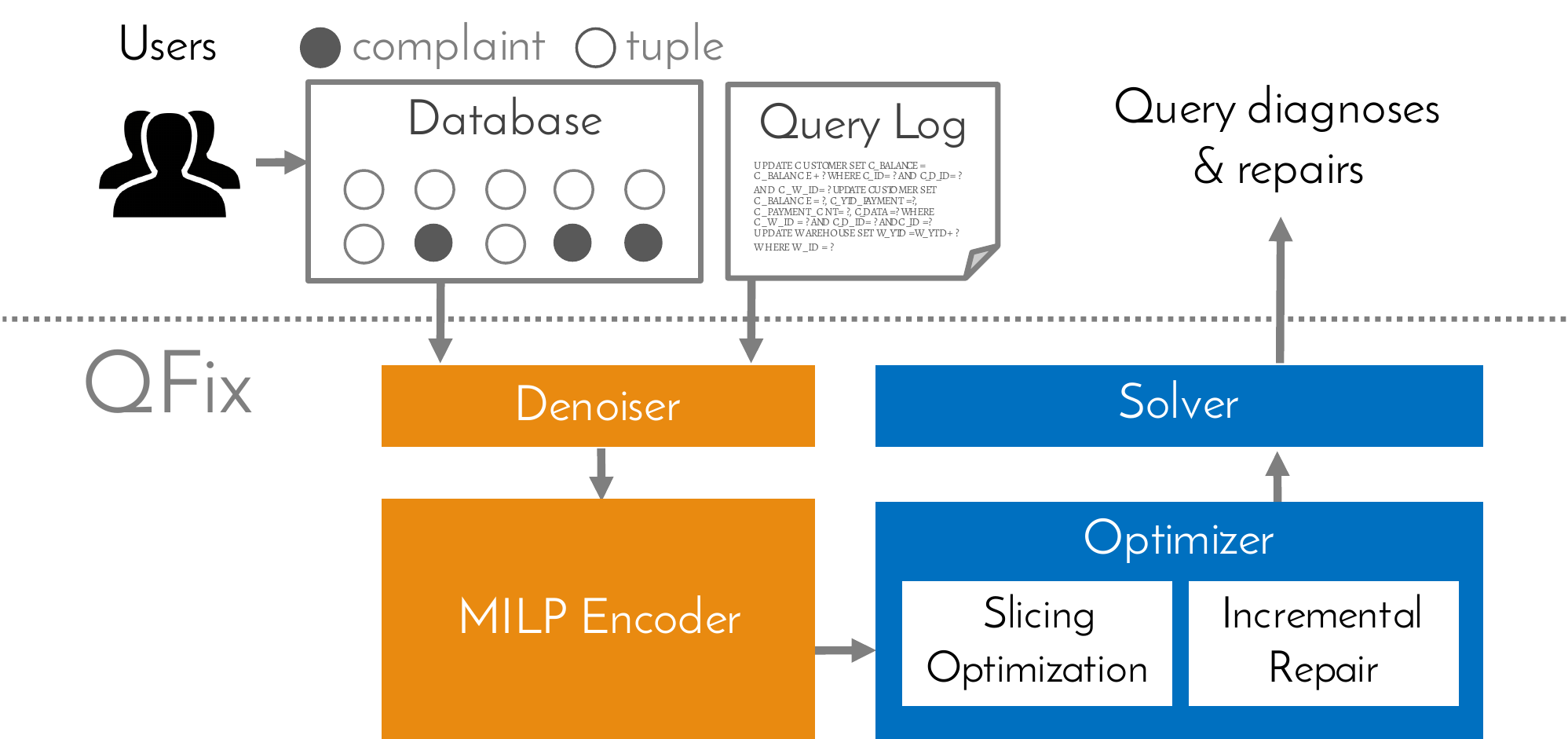}
    \caption{\sys processes data anomalies in the form of complaints and analyzes logged query histories to identify the causes of error in the form of repaired queries. 
    The key component is the MILP Encoder, which expresses the diagnosis problem as a mixed integer linear program.}
    \label{fig:architecture}
\end{figure}

Figure~\ref{fig:architecture} shows the \sys architecture. The system takes
two inputs: a log of update queries (including \texttt{UPDATE}, \texttt{INSERT}, and \texttt{DELETE}
statements) and a set of identified data errors (\emph{complaints}). \sys
analyzes the data errors and the query logs to trace the causes of the errors
in queries in the log (diagnoses), and to automatically derive query repairs.
The query repairs represent corrections to the queries in the log, and can be
used to identify additional errors in the data that were not reported.
 
The core of \sys is the \emph{MILP Encoder}, which expresses the query diagnosis
problem as a Mixed Integer Linear Program (MILP), and the constraint problem is solved by the \emph{Solver}. 
The \emph{Optimizer} uses slicing and incremental techniques that
help the system scale to large datasets and query logs efficiently, while maintaining high accuracy. 
For completeness, an optional \emph{Denoiser} can be applied to the inputs to detect and remove incorrect complaints---we regard this component as orthogonal to this paper.

\section{Modeling abstractions}
\label{sec:abstractions}

In this section, we introduce a running example inspired from the use-case of
Example~\ref{ex:taxes}, and describe the model abstractions that we use to
formalize the diagnosis problem.

\begin{figure*}[t]
    \begin{minipage}[t]{0.28\textwidth}
         \vspace{0pt} 
         \centering
        \begin{tabular}{llll}
            \multicolumn{4}{l}{\emph{Taxes}: $D_0$}\\
            \toprule
            \textbf{ID}  & \textbf{income}    & \textbf{owed} & \textbf{pay} \\
            \midrule
            $t_1$   & \$9500    & \$950		& \$8550 \\
            $t_2$   & \$90000   & \$22500 	& \$67500\\
            $t_3$   & \$86000   & \$21500	& \$64500\\
            $t_4$   & \$86500   & \$21625	& \$64875\\
            \bottomrule
            \\
        \end{tabular}
    \end{minipage}
    \begin{minipage}[t]{0.43\textwidth}
         \vspace{0pt} 
         \centering
        \begin{tabular}{|p{1ex}l|}
            \multicolumn{2}{l}{\emph{Query log}: $\mathcal{Q}$}\\
            \hline
            
            $q_1$: & \texttt{\small UPDATE Taxes SET owed=income*0.3}\\
            	   & \texttt{\small WHERE income>=\color{red}{85700}}\\
            
            $q_2$: & \texttt{\small INSERT INTO Taxes}\\ 
                   & \texttt{\small VALUES (25, 85800, 21450)}\\
                   
            $q_3$: & \texttt{\small UPDATE Taxes SET pay=income-owed}\\ 
            \hline
        \end{tabular}
    \end{minipage}
    \begin{minipage}[t]{0.28\textwidth}
         \vspace{0pt} 
         \centering
        \begin{tabular}{llll}
            \multicolumn{4}{l}{\emph{Taxes}: $D_4$}\\
            \toprule
            \textbf{ID}  & \textbf{income}    & \textbf{owed}  & \textbf{pay}\\
            \midrule
            $t_1$ & \$9500    & \$950		& \$8550\\
            $t_2$   & \$90000   & \$27000	& \$63000\\
            \rowcolor{mid-gray}
            $t_3$ & \$86000   & \color{red}{\$25800} & \color{red}{\$60200}\\
            \rowcolor{mid-gray}
            $t_4$	 & \$86500	  & \color{red}{\$25950}	& \color{red}{\$60550}\\
            $t_5$	 & \$87000	  & \$21750	& \$65250\\
            \bottomrule
        \end{tabular}
    \end{minipage}

    \vspace{-2mm}
    \caption{A recent change in tax rate brackets calls for a tax rate of 30\% for those with income above \$87500.  The accounting department issues query $q_1$ to implement the new policy, but the predicate of the WHERE clause condition transposed two digits of the income value.
\vspace*{-0.1in}
    }
    \label{fig:example}
\end{figure*}

\begin{example}\label{ex:taxes2}
Figure~\ref{fig:example} demonstrates an example tax bracket adjustment in the
spirit of Example~\ref{ex:taxes}. The adjustment sets the tax rate to 30\% for
income levels above \$87,500, and is implemented by query $q_1$. A digit
transposition mistake in the query, results in an incorrect owed amount for tuples
$t_3$ and $t_4$. Query $q_2$, which inserts a tuple with slightly higher
income than $t_3$ and $t_4$ and the correct information, obscures this mistake.
This mistake is further propagated by query $q_3$, which calculates the pay 
check amount based on the corresponding
income and owed. 
\end{example}
\vspace*{-0.07in}
While traditional data cleaning techniques seek to identify and correct the
erroneous values in the table \emph{Taxes} directly, our goal is to diagnose
the problem, and understand the reasons for these errors. In this case, the
reason for the data errors is the incorrect predicate value in query $q_1$.

In this paper, we assume that we know \emph{some} errors in the dataset, and
that these errors were caused by erroneous updates. The errors may be
obtained in different ways: traditional data cleaning tools may identify
discrepancies in the data (e.g., a tuple with lower income has higher owed tax
amount), or errors can be reported directly from users (e.g., customers
reporting discrepancies to customer service). \emph{Our goal is not to correct
the errors directly in the data, but to analyze them as a ``symptom'' and provide a
diagnosis.} The diagnosis can produce a targeted treatment: knowing how the
errors were introduced guides the proper way to trace and resolve them.

\begin{figure}[t]
\centering
{\small
\begin{tabular}{ll}
    \toprule
    \textbf{Notation} & \textbf{Description}\\
    \midrule
    $\mathcal{Q}$& The sequence of executed update queries (log)\\ 
             & $\mathcal{Q}=\{q_1, \dots, q_n\}$ \\
    $D_0$    & Initial database state at beginning of log\\
    $D_n$    & End database state (current) $D_n=\mathcal{Q}(D_0)$\\
    $D_i$    & Database state after query $q_i$: $D_i=q_i(\dots q_1(D_0))$\\
    $c: t\mapsto t^*$ & Complaint: $\mathcal{T}_c(D) = (D_n\setminus\{t\})\cup\{t^*\}$\\
    $\mathcal{C}$ & Complaint set $\mathcal{C}=\{c_1,\dots,c_k\}$\\
    $\mu_q(t)$  & Modifier function of $q$ (e.g., \texttt{SET} clause)\\
    $\sigma_q(t)$   & Conditional function of $q$ (e.g., \texttt{WHERE} clause)\\
    $t_{new}$   & Tuple values introduced in an \texttt{INSERT} query\\
    $\mathcal{Q}^*$& Log repair\\
    $d(\mathcal{Q}, \mathcal{Q}^*)$ & Distance functions between two query logs\\
    \bottomrule
\end{tabular}
}
\vspace{-2mm}
\caption{Summary of notations used in the paper. }
\label{tbl:notation}
\end{figure}

\subsection{Error Modeling}
\label{sec:model}

In our setting, the diagnoses are associated with errors in the queries that
operated on the data. In Example~\ref{ex:taxes2}, the errors in the dataset
are due to the digit transposition mistake in the WHERE clause predicate of
query $q_1$. Our goal is to infer the errors in a log of queries
automatically, given a set of incorrect values in the data. We proceed to
describe our modeling abstractions for data, queries, and errors, and how we
use them to define the diagnosis problem.

\subsubsection*{Data and query models}
\label{sec:models}

\noindent
\emph{Query log ($\mathcal{Q}$):}
We define a query log that update the database 
as an ordered sequence of \texttt{UPDATE}, \texttt{INSERT}, and
\texttt{DELETE} queries $\mathcal{Q}=\{q_1,\dots,q_n\}$, that have
operated on a database $D$. In the rest of the paper, we use the term
\emph{update queries}, or just \emph{queries}, to refer to any of the queries in $\mathcal(Q)$,
including insertion and deletion queries.

\smallskip
\noindent
\emph{Query ($q_i$):} We model each query as a function over a
database $D$, resulting in a new database $D'$. For \texttt{INSERT}
queries, $D'=q(D)=D\cup\{t_{new}\}$.
We model \texttt{UPDATE} and \texttt{DELETE} queries as follows:  
\begin{align*}
    D'=q(D)= &\{\mu_{q}(t)\;|\;t\in D, \sigma_{q}(t)\}
    \cup\{t\;|\;t\in D, \neg\sigma_{q}(t)\}
\end{align*}
In this definition, the modifier function $\mu_q(t)$ represents the query's update equations, and it transforms a tuple by either deleting it ($\mu_q(t)=\bot$) or changing the values of some of its attributes.
The conditional function $\sigma_q(t)$ is a boolean function that represents the query's condition predicates.  In the example of Figure~\ref{fig:example}:
\begin{align*}
    &\mu_{q_1}(t)=(t.income, t.income*0.3, t.pay)\\
    &\sigma_{q_1}(t)=(t.income\ge 85700)\\
    &\mu_{q_3}(t)=(t.income, t.owed, t.income-t.owed)\\
    &\sigma_{q_2}(t)=\texttt{true}
\end{align*} 
Note that in this paper, we only consider query without sub-query or aggregation.

\smallskip
\noindent
\emph{Database state ($D_i$):}
We use $D_i$ to represent the state of a database $D$ after the application of
queries $q_1$ through $q_i$ from the log $\mathcal{Q}$. $D_0$ represents the
original database state, and $D_n$ the final, or current, database state. Out
of all the states, the system only maintains $D_0$ and $D_n$. In practice,
$D_0$ can be a checkpoint: a state of the database that we assume is correct;
we cannot diagnose errors before this state. The intermediate states can be
derived by executing the log: $D_i=q_i(q_{i-1}(\dots q_1(D_0)))$. We also
write $D_n=\mathcal{Q}(D_0)$ to denote that the final database state $D_n$ can
be derived by applying the sequence of queries in the log to the original
database state $D_0$.

\smallskip
\noindent
\emph{True database state ($D_i^*$):}
Queries in $\mathcal{Q}$ are possibly erroneous, introducing errors in the
data. There exists a sequence of \emph{true} database states $\{D_0^*,
D_1^*\dots, D_n^*\}$, with $D_0^*=D_0$, representing the database states that
would have occurred if there had been no errors in the queries.
The true database states are unknown; our goal is to find and correct the errors in $\mathcal{Q}$ and retrieve the correct database state $D_n^*$.

For ease of exposition, in the remainder of the paper we assume that the
database contains a single relation with attributes $A_1,\ldots,A_m$,
but the single table is not a requirement in our framework.

\subsubsection*{Error models}

Following the terminology in Examples~\ref{ex:telco}
and~\ref{ex:taxes}, we model a set of identified or user-reported
data errors as \emph{complaints}. A complaint corresponds to a
particular tuple in the final database state $D_n^*$, and identifies
that tuple's correct value assignment. We formally define complaints
below:
\begin{definition}[Complaint]
    A complaint $c$ is a mapping between two tuples: $c: t\mapsto t^*$, such that $t$ and $t^*$ have the same schema, $t\in D_n\cup\{\bot\}$, and $t\neq t^*$. A complaint defines a
    transformation $\mathcal{T}_c$ on a database state $D$: $\mathcal{T}_c(D)
    = (D\setminus\{t\})\cup\{t^*\}$.
\end{definition}

In the example of Figure~\ref{fig:example}, two complaints are reported on the final database state $D_3$: 
$c_1: t_3\mapsto t_3^*$ and
$c_2: t_4\mapsto t_4^*$, 
where $t_3^*=(86000,21500,64500)$ and $t_4^*=(86500,21625,\linebreak 64875)$. 
For both these cases, each complaint denotes a \textbf{value correction} for a tuple in $D_3$.  Complaints can also model the \textbf{addition} or \textbf{removal} of tuples: $c: \bot\mapsto t^*$ means that $t^*$ should be added to the database, whereas $c: t\mapsto \bot$
means that $t$ should be removed from the database.

\smallskip
\noindent
\emph{Complaint set ($\mathcal{C}$):}
We use $\mathcal{C}$ to denote the set of all known complaints
$\mathcal{C}=\{c_1,\dots,c_k\}$, and we call it the \emph{complaint set}.
Each complaint in $\mathcal{C}$ represents a transformation (addition,
deletion, or modification) of a tuple in $D_n$. We assume that the
complaint set is consistent, i.e., there are no two complaints that
propose different transformations to the same tuple $t\in D_n$.
Applying all these transformations to $D_n$ results in a new database
instance
$D_n'=\mathcal{T}_{c_1}(\mathcal{T}_{c_2}(\dots\mathcal{T}_{c_k}(D_n)))$.\footnote{Since
the complaint set is consistent, it is easy to see that the order of
transformations is inconsequential.} $\mathcal{C}$ is \emph{complete}
if it contains a complaint for each error in $D_n$. In that case,
$D_n'=D_n^*$. In our work, we do not assume that the complaint set is
complete, but, as is more common in practice, we only know a subset of
the errors (incomplete complaint set). Further, we focus our analysis
on \emph{valid} complaints; we briefly discuss dealing with invalid
complaints (complaints identifying a correct value as an error) in
Section~\ref{sec:noise}, but these techniques are beyond the scope of this paper.

\smallskip
\noindent
\emph{Log repair ($\mathcal{Q}^*$):}
The goal of our framework is to derive a diagnosis as a log repair
$\mathcal{Q}^*=\{q_1^*,\dots, q_n^*\}$, such that
$\mathcal{Q}^*(D_0)=D_n^*$. In this work, we focus on errors produced
by incorrect parameters in queries, so our repairs focus on altering
query constants rather than query structure. Therefore, for each query
$q_i^*\in\mathcal{Q}^*$, $q_i^*$ has the same structure as $q_i$
(e.g., the same number of predicates and the same variables in the \texttt{WHERE} clause), 
but possibly different parameters. For example, a good log repair for the
example of Figure~\ref{fig:example} is
$\mathcal{Q}^*=\{q_1^*,q_2,q_3\}$, where $q_1^*$=\texttt{UPDATE Taxes
SET owed=income*0.3 WHERE income >= 87500}.

\subsubsection*{Problem definition}

We now formalize the problem definition for diagnosing data
errors using query logs. A diagnosis is a log repair
$\mathcal{Q}^*$ that resolves all complaints in the set $\mathcal{C}$
and leads to a correct database state $D_n^*$.
\begin{definition}[Optimal diagnosis]\label{def:problem}
    Given database states $D_0$ and $D_n$, a query log $\mathcal{Q}$ such that $\mathcal{Q}(D_0)=D_n$, a set of complaints $\mathcal{C}$ on $D_n$,  and a distance function $d$, the optimal diagnosis is a log repair $\mathcal{Q}^*$, such that:
    \begin{itemize}[itemsep=0pt, parsep=0pt, topsep=1pt]
        \item $\mathcal{Q}^*(D_0)=D_n^*$, where $D_n^*$ has no errors
        \item $d(\mathcal{Q}, \mathcal{Q}^*)$ is minimized
    \end{itemize}
\end{definition}

More informally, we seek the minimum changes to the log $\mathcal{Q}$
that would result in a clean database state $D_n^*$. Obviously, a
challenge is that $D_n^*$ is unknown, unless we know that the
complaint set is complete. 

\subsubsection*{Problem scope and solution outline}
In this work, we assume basic data manipulation queries with no
subqueries, aggregations, or joins; these operations are not as common in update workloads. \sys supports queries with WHERE
clauses containing conjunctions and disjunctions of predicates.
Predicates and SET expressions can be over linear combinations of
constants and attributes. We study the impact of the number of
predicates in the WHERE clause in Section~\ref{sec:experiments:synth}.

In Section~\ref{sec:sol}, we describe our basic method, 
which
uses a constraint programming formulation that expresses this
diagnosis problem as a mixed integer linear program (MILP). 
Section~\ref{sec:opt} presents several optimization
techniques that extend the basic method, allowing \sys to 
(1)~handle cases of incomplete information (incomplete complaint set), and
(2)~scale to large data and log sizes. 
Specifically, the fully optimized, incremental algorithm (Section~\ref{sec:incremental}), can
handle query logs with hundreds of queries within minutes, while the performance of the basic approach collapses by $50$ queries.

Due to space considerations, we omit discussion of alternative approaches that use classification tools and linear systems of equations.
These approaches are limited to a query log containing a single query, and are discussed and evaluated in more detail in our technical report~\cite{qfixarxiv}.

\section{A MILP-based Solution}
\label{sec:sol}

In this section, we introduce a \emph{basic} solver-based approach to 
resolve the errors reflected in the complaint set.
This approach constructs a mixed-integer linear 
programming (MILP) problem by linearizing and parameterizing the 
corrupted query log over the tuples in the database. 
Briefly, an MILP is a linear program where only a subset of the undetermined variables
are required to be integers, while the rest are real-valued.

\looseness -1
Our general strategy is to model each query as a linear equation 
that computes the output tuple values from the inputs and to transform the
equation into a set of of linear constraints.   
In addition, the constant values in the queries are parameterized
into a set of undetermined variables, while the database state is encoded 
as constraints on the initial and final tuple values.
Finally, the undetermined variables are used to construct an objective function
that prefers value assignments that minimize both the amount that the queries change and
the number of non-complaint tuples that are affected. 

The rest of this section will first describe the process of linearizing a single query
and translating it into a set of constraints.  We then extend the process to the entire
query log and finally define the objective function.
Subsequent sections introduce optimizations that both
improve the speed and quality of the results, as well as harness the trade-off between the two.

\subsection{Encoding a Single Query}
\label{sec:linearize}

\if{0}
  We model a query $q_i$ as a conditional function $f_{q_i}(t)$ that takes as input a tuple $t$
  and returns its next state $t'$.  $f_{q_i}$ is applied to each 
  tuple $t \in \mathcal{D}_{i-1} \cup \{\bcancel{t}\}$ in the input relation along with a special
  non-existant tuple $\bcancel{t}$. \ewu{maybe fold this into the data model.}
  By treating the query as a function, we are able to encode its effects into a set
  of linear inequality constraints.  We call this process the linearization and 
  parameterization of a query.

  \begin{definition} [Conditional Function]
  \label{def:cond}
    The conditional function for query $q$ is:
    \[
      f_{q_i}(t)= 
      \begin{cases}
      f_{q_i.\mu} (t) ,& \text{if } f_{q_i.\sigma} (t)\\
      t,              & \text{otherwise}
      \end{cases}
  \]
  where the \textit{update function} $f_{q_i.\mu}$ models a set of \textit{update equation(s)};
  and the \textit{condition function} $f_{q_i.\sigma}$ models a set of \textit{logical expression(s)} in 
  disjunctive or conjunctive form.
  \end{definition}

  Conditional functions can describe the common classes of update queries:
  \begin{enumerate}
  \item \texttt{UPDATE}: $f_{q_i.\mu}(t)$, $f_{q_i.\sigma}(t)$ model the \texttt{SET}
        and \texttt{WHERE} clauses.  For example, $f_{q_i.\mu}(<t.a, t.b>) = <t.a + 1, 2>$ and
        $f_{q_i.\sigma}(t) = (t.a > 20)$ corresponds to the query 
        \texttt{UPDATE D SET a = a + 1, b = 2 WHERE a > 20}.

  \item \texttt{INSERT}: $f_{q_i.\mu}(t)$ returns the inserted tuple, while 
        $f_{q_i.\sigma}(t) = (t = \bcancel{t})$ evaluates to true when it is executed over
        the special nonexistant tuple.

  \item \texttt{DELETE}: $f_{q_i.\mu}(t) = \bcancel{t}$ returns a nonexistant tuple whenever
        the predicate encoded in $f_{q_i.\sigma}(t)$ evaluates to true.
        For example, the query \texttt{DELETE FROM T WHERE a < 20} represents 
        $f_{q_i.\sigma}(t) = (t.a < 20)$.
        
  \end{enumerate}

  Finally, we parameterize $q_i$ by replacing all numeric constants in the
  conditional function with undetermined variables.   Consider the conditional
  function for \texttt{UPDATE} above: the constants $1$, $2$ in $f_{q_i.\mu}$
  as well as $20$ in $f_{q_i.\sigma}$ will be transformed into undetermined
  variables \texttt{v1, v2, v3} that are solved by the MILP solver.

\fi

MILP problems express constraints as a set of linear inequalities. Our
task is to derive such a mathematical representation for each query in
$\mathcal{Q}$. Starting with the functional representation of a
query (Section~\ref{sec:models}), we describe how each query type,
\texttt{UPDATE}, \texttt{INSERT}, and \texttt{DELETE}, can be
transformed into a set of linear constraints over a tuple $t$ and an
attribute value $A_j$.

\smallskip
\noindent
\texttt{UPDATE:}
Recall from Section~\ref{sec:models} that query $q_i$ can be modeled
as the combination of a modifier function $\mu_{q_i}(t)$ and
conditional function $\sigma_{q_i}(t)$. First, we introduce a binary
variable $x_{q_i, t}$ to indicate whether $t$ satisfies the
conditional function of $q_i$: $x_{q_i, t}=1$ if
$\sigma_{q_i}(t)=\texttt{true}$ and $x_{q_i, t}=0$ otherwise. In a
slight abuse of notation:
\begin{align}
\label{eq:x}
x_{q_i, t} = \sigma_{q_i}(t)
\end{align}
Next, we introduce real-valued variables for the attributes of $t$.
We express the updated value
of an attribute using semi-modules, borrowing from the models of
provenance for aggregate operations~\cite{Amsterdamer2011}. A
semi-module consists of a commutative semi-ring, whose elements are
scalars, a commutative monoid whose elements are vectors, and a
multiplication-by-scalars operation that takes a scalar $x$ and a
vector $u$ and returns a vector $x \otimes u$. A similar formalism has
been used in literature to model hypothetical data
updates~\cite{tiresias}.

Given a query $q_i$ and tuple $t$, we express the value of attribute $A_j$ in the updated tuple $t'$ as follows:
\begin{align}
\label{eq:linearization}
t'.A_j = x_{q_i, t}\otimes \mu_{q_i}(t).A_j + (1-x_{q_i, t})\otimes t.A_j 
\end{align} 
\looseness -1
In this expression, the $\otimes$ operation corresponds to regular
multiplication, but we maintain the $\otimes$ notation to indicate
that it is a semi-module multiplication by scalars. This expression
models the action of the update: If $t$ satisfies the conditional
function ($x_{q_i, t}=1$), then $t'.A_j$ takes the value
$\mu_{q_i}(t).A_j$; if $t$ does not satisfy the conditional function
($x_{q_i, t}=0$), then $t'.A_j$ takes the value $t.A_j$.
In our running example, the rate value of a tuple $t$ after query $q_1$ would be expressed as:
$t'.owed = x_{q_1, t}\otimes (t.income*0.3) + (1-x_{q_1, t})\otimes t.owed$.
Equation~\eqref{eq:linearization} does not yet provide a linear
representation of the corresponding constraint, as it contains
multiplication of variables. To linearize this expression, we adapt a
method from~\cite{tiresias}: We introduce two variables $u.A_j$ and
$v.A_j$ to represent the two terms of
Equation~\eqref{eq:linearization}: $u.A_j=x_{q_i, t}\otimes
\mu_{q_i}(t).A_j$ and $v.A_j=(1-x_{q_i, t})\otimes t.A_j$. Assuming a
number $M$ as the upper bound of the domain of $t.A_j$, we get the
following constraints:
\begin{align}
\label{eq:uv}
u.A_j &\!\leq\! \mu_{q_i}(t).A_j   &v.A_j &\!\leq\! t.A_j &\nonumber\\
u.A_j &\!\leq\! x_{q_i, t}M        &v.A_j &\!\leq\! (1\!-\!x_{q_i, t})M &\\
u.A_j &\!\geq\! \mu_{q_i}(t).A_j \!-\! (1\!-\! x_{q_i, t})M \phantom{i} 
&v.A_j &\!\geq\! t.A_j \!-\! x_{q_i, t}M &\nonumber
\end{align}
The set of conditions on $u.A_j$ ensure that $u.A_j = \mu_{q_i}(t).A_j$ if $x_{q_i, t}=1$, and $0$ otherwise. Similarly, 
the conditions on $v.A_j$ ensure that $v.A_j = t.A_j$ if $x_{q_i, t}=0$, and $0$ otherwise.  
Now, Equation~\eqref{eq:linearization} becomes linear:
\begin{align}
\label{eq:tnew}
t.A_j' = u.A_j + v.A_j
\end{align}

\noindent\texttt{INSERT:}\looseness -1
An insert query adds a new tuple $t_{new}$ to the database.  If the query were 
corrupted, then the inserted values need repair. We use a binary variable $x$ to model whether the query is correct.  Each attribute of the newly inserted tuple ($t'.A_j$) may take one of two values: the value specified by the insertion query ($t_{new}.A_j$) if the query is correct ($x=1$), or an undetermined value ($u.A_j$) if the query is incorrect ($x=0$).  Thus, similar with Equation~\eqref{eq:linearization}, we write:
\begin{eqnarray}
\label{eq:insert}
t'.A_j = x \otimes t_{new}.A_j + (1-x) \otimes v.A_j 
\end{eqnarray}

\smallskip
\noindent
\texttt{DELETE:}
A delete query removes a set of tuples from the database.  
Since the MILP problem doesn't have a way to express a non-existent value, 
we encode a deleted tuple by setting its attributes to a value
outside of the attribute domain $M^+$.  In this way, subsequent conditional functions
on the attribute will return false, so it will not have an effect on subsequent queries encoded
in the MILP problem:
\begin{eqnarray}
\label{eq:delete}
t'.A_j &=& x_{q_i, t} \otimes M^+ + (1-x_{q_i, t}) \otimes t.A_j \\
x_{q_i, t} &=& \sigma_{q_i}(t)\nonumber 
\end{eqnarray}
This expression is further linearized using the same method as Equation~\eqref{eq:uv}.

\smallskip
\noindent
\textbf{Putting it all together.}
The constraints defined in Equations \eqref{eq:x}--\eqref{eq:delete}
form the main structure of the MILP problem for a single attribute
$A_j$ of a single tuple $t$. To linearize a query $q_i$ one needs to
apply this procedure to all attributes and tuples. This process is
denoted as $Linearize(q, t)$ in Algorithm~\ref{alg:basic}. Our MILP
formulation includes three types of variables: the binary variables
$x_{q_i, t}$, the real-valued attribute values (e.g., $u.A_j$), and
the real-valued constants in $\mu_{q_i}$ and $\sigma_{q_i}$. All these
variables are undetermined and need to be assigned values by a MILP
solver.

Next, we extend this encoding to the entire query log,
and incorporate an objective function encouraging solutions
that minimize the overall changes to the query log.

\subsection{Encoding and Repairing the Query Log}
\label{sec:milp}

We proceed to describe the procedure (Algorithm~\ref{alg:basic}) that encodes 
the full query log into a MILP problem, and solves the MILP problem to derive $\mathcal{Q}^*$.
The algorithm takes as input the query log $\mathcal{Q}$, 
the initial and final (dirty) database states 
$\mathcal{D}_{0, n}$, and the complaint set $\mathcal{C}$, and outputs a fixed query 
log $\mathcal{Q}^*$.  

\looseness -1
We first call \textit{Linearize} on each tuple in $\mathcal{D}_0$ and
each query in $\mathcal{Q}$, and add the result to a set of
constraints \textit{milp\_cons}. The function \textit{AssignVals} adds
constraints to set the values of the inputs to $q_0$ and the outputs
of $q_n$ to their respective values in $\mathcal{D}_0$ and
$\mathcal{T}_\mathcal{C}(\mathcal{D}_n)$. Additional constraints
account for the fact that the output of query $q_i$ is the input of
$q_{i+1}$ (\textit{ConnectQueries}). This function simply equates $t'$
from the linearized result for $q_i$ to the $t$ input for the
linearized result of $q_{i+1}$.

Finally, \emph{EncodeObjective} augments the program with an objective
function that models the distance function between the original query
log and the log repair ($d(\mathcal{Q},\mathcal{Q}^*)$). In the
following section we describe our model for the distance function,
though other models are also possible. Once the MILP solver returns a
variable assignment, \textit{ConvertQLog} updates the constants in the
query log based on this assignment, and constructs the fixed query log
$\mathcal{Q}^*$.

\begin{algorithm}[t]
\caption{$Basic:$ The MILP-based approach.}
\label{alg:basic}
\scriptsize
\begin{algorithmic}[1]
\REQUIRE {$\mathcal{Q}, D_0, D_n, \mathcal{C}$}
\STATE $milp\_cons \leftarrow \emptyset$
\FOR {each $t$ in $R$}
\FOR {each $q$ in $\mathcal{Q}$}
\STATE $milp\_cons \leftarrow milp\_cons \cup Linearize(q, t)$
\ENDFOR
\STATE $milp\_cons \leftarrow milp\_cons \cup AssignVals(D_0.t, D_n.t, \mathcal{C})$
\FOR {each $i$ in $\{0,\ldots,N-1\}$}
\STATE $milp\_cons \leftarrow milp\_cons \cup ConnectQueries(q_i, q_{i+1})$
\ENDFOR
\ENDFOR 
\STATE $milp\_obj \leftarrow EncodeObjective(milp\_cons, \mathcal{Q})$
\STATE $solved\_vals \leftarrow MILPSolver(milp\_cons, milp\_obj)$
\STATE $\mathcal{Q}^* \leftarrow ConvertQLog(Q, solved\_vals)$
\STATE Return $\mathcal{Q}^*$
\end{algorithmic}
\end{algorithm}

\subsection{The Objective Function}

The optimal diagnosis problem (Definition~\ref{def:problem}) seeks a
log repair $\mathcal{Q}^*$, such that the distance
$d(\mathcal{Q},\mathcal{Q}^*)$ is minimized. In this section, we
describe our model for the objective function, which assumes numerical
parameters and attributes. This assumption is not a restriction of the
\sys framework.
Handling other data types, such as categorical values comes down to defining an appropriate distance function, which can then be directly incorporated into \sys.

In our experiments, we use the normalized Manhattan
distance (in linearized format in the MILP problem) 
between the parameters in $\mathcal{Q}$ and
$\mathcal{Q}^*$. We use $q.param_i$ to denote the $i^{th}$ parameter
of query $q$, and $|q.param|$ to denote the total number of parameters
in $q$: \[d(\mathcal{Q}, \mathcal{Q}^*) = \sum_{i = 1} ^{n} \sum_{j =
1}^{|q_i.param|} |q_i.param_j - q_i.param_j^*|\]
Different choices for the objective function are also possible. For
example, one may prioritize the total number of changes incurred in
the log, rather than the magnitude of these changes. However, a
thorough investigation of different possible distance metrics is
beyond the scope of our work.

\section{Optimizing the Basic Approach}
\label{sec:opt}

\begin{figure}[t]
    \centering
    \includegraphics[width=0.35\textwidth]{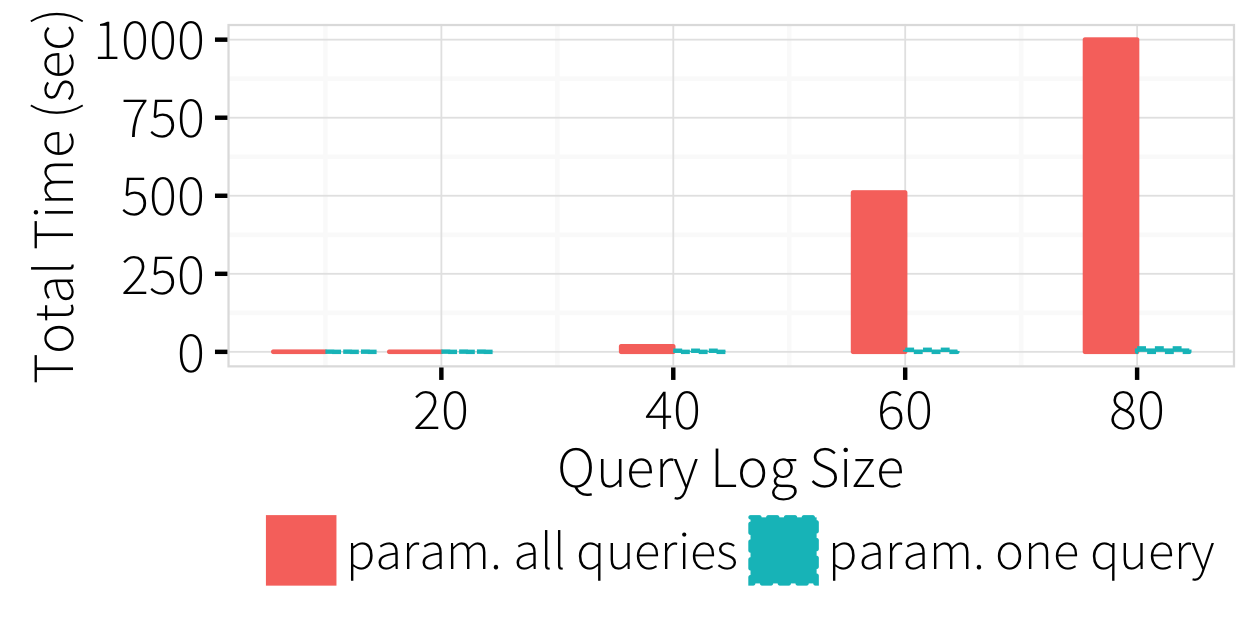}
    \vspace*{-0.1in}
    \caption{Log size vs. execution time over 1000 records. }
    \label{fig:querysize_vs_time}
\end{figure}

A major drawback of our \emph{basic} MILP transformation (Section~\ref{sec:sol}) is
that it exhaustively encodes the combination of all tuples in the database and all queries
in the query log.  In this approach, the number of constraints (as well as undetermined variables) 
grows quadratically with respect to the database and the query log.
This increase has a large impact on the running time of the solver, since it needs to find a (near)-optimal 
assignment of all undetermined variables (exponential with the number of undetermined variables).
This is depicted in Figure~\ref{fig:querysize_vs_time}, which increases the query log size over a database of $1000$ tuples.  
The red bars encode the problem using  the \emph{basic} algorithm that parameterizes all queries, while the blue bars show the potential gain of only parameterising the oldest query that we assume is incorrect.
At a log size of $80$, the solver for \emph{basic} failed to produce an answer within $1000$ seconds.
Although MILP solvers exhibit empirical performance variation,
this experiment illustrates the performance limitation of the \emph{basic} approach. 

A second limitation of \emph{basic} is its inability to handle errors in the complaint set.
This is because the \emph{basic} MILP formulation generates hard constraints for all of the database records, thus any error, whether a false negative missing complaint or a false positive incorrect complaint, must be correct.
It may be impossible to find a repair that satisfies this condition and will lead to solver infeasibility errors.

The rest of this section describes three classes of \emph{slicing} optimizations that 
reduce the number of tuples, queries, and attributes that are encoded in the MILP problem. 
The tuple-slicing technique additionally improves the repair accuracy when the complaint set is incomplete. 
We also propose an incremental algorithm that avoids the exponential increase in solver time by only parameterizing a small number of queries at a time---thus limiting
the cost to the left side of Figure~\ref{fig:querysize_vs_time}.

\begin{figure}[t]
    \centering
    \includegraphics[width=0.35\textwidth]{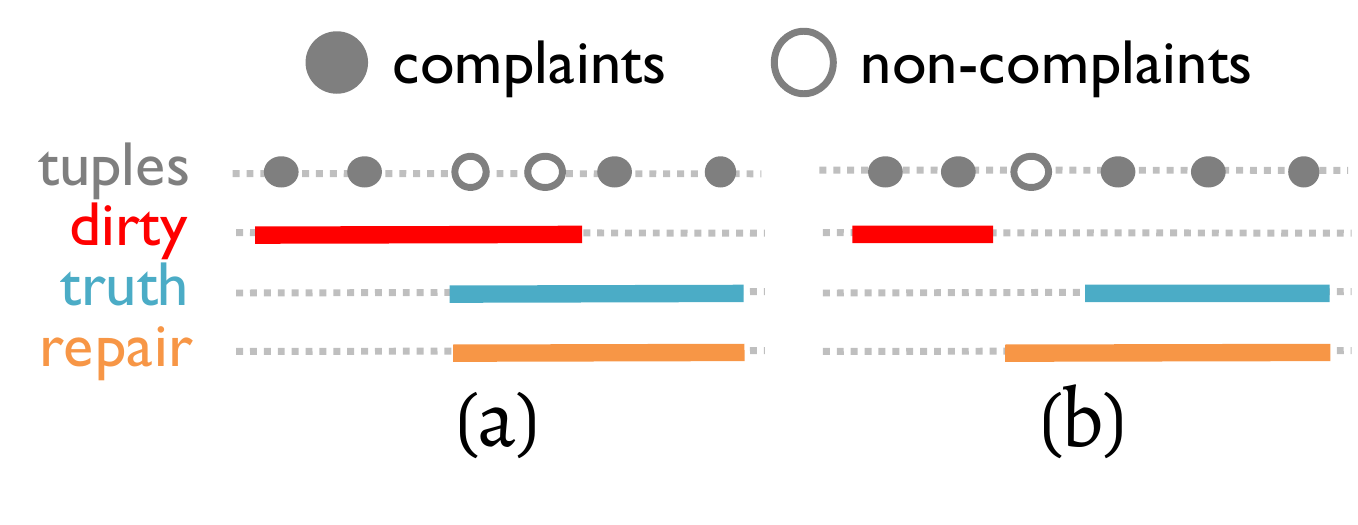}
    \vspace*{-2mm}
    \caption{
      Graphical depiction of correct (a) and over-generalized (b) repairs.
      Solid and empty circles represent complaint and non-complaint tuples.
      Each thick line represents the interval of query $q$'s range predicate.
      Dirty: incorrect interval in corrupted query;
      truth: correct interval in true query;
      repair: interval returned by the solver.}
    \label{fig:groups}
\end{figure}

\if{0}
linearize whole query log, so cost of adding an additional tuple is very high.
second iteration generally takes ~1 - 10
for large databases knn cost is pretty high: ~

if the solver returns, it is always a super set of the clean range

only tuples modified by the fixed queries
- tuples already in the complaints (correct)
- not in complaints, but any of the originial queries modified it
- not in complaints, but no original queries modified it
\fi

\subsection{Tuple Slicing: Reducing Tuples}
\label{sec:opt:tbsize}

Our first optimization, \emph{tuple-slicing}, applies a two step process to reduce the 
problem size without sacrificing accuracy: it
first aggressively reduces
the problem size by only encoding tuples in the complaint set and then refines the log repair 
through a second but much smaller MILP problem. 

\smallskip
\noindent\textbf{Step 1 (Initial Repair Step):}
The first step of  \emph{tuple slicing} 
aggressively reduces the problem size by only encoding those 
tuples in the complaint set $\mathcal{C}$ (Algorithm~\ref{alg:basic} line $2$
is replaced with \texttt{for each t in $\mathcal{C}$}). 
Each tuple necessitates
the linearization of the entire query log, thus, only encoding the complaint tuples minimizes the 
size of the problem with respect to the relevant tuples. 
This optimization is guaranteed to resolve $\mathcal{C}$, thus depending on the properties of the non-complaint records, it can generate correct repairs
an order of magnitude faster without hurting the accuracy. 
In Figure~\ref{fig:groups}(a),
the solver will guarantee a repair interval that excludes the two left-most complaints, includes the two right-most complaints,
and minimizes the difference between the dirty and repaired intervals (due to the objective function).
This effectively pushes the repair's lower-bound towards that of the dirty interval.
This is a case where such a solution is correct, because the dirty and truth intervals overlap.
Recall that we do not have access to the truth interval, and our goal is to reproduce the truth interval given $\mathcal{C}$ (solid circles) and the corrupted query.

However, this approach can also cause the repair to be a {\it superset} of the truth interval, and affect tuples not part of the complaint set.
Figure~\ref{fig:groups}(b) highlights such a case where the dirty and truth intervals are non-overlapping, and the non-complaint record between them has been incorrectly included in the repair interval---\emph{because the MILP problem did not include the non-complaint.}

\indent In both of these cases, the objective function will ensure that the repair
does not over-generalize the upper bound towards the right because that strictly increases the objective function.
Therefore, our main concern is to refine the repair interval to exclude those non-complaint tuples in case (b).
Note that in the case of incomplete complaint sets, the user may opt to no execute the refinement step if she believes
that the non-complaint records are indeed in error.

\smallskip

\noindent\textbf{Step 2 (Refinement Step):} \looseness -1
Although there are many possible mechanisms to refine the initial repair (e.g., incrementally shrinking
the repaired interval until the non-complaint tuples are all excluded), 
the straightforward approaches are not effective when multiple corrupt 
queries have been repaired because they don't take the query interactions into account.

\looseness -1
Instead, we solve this with a second, significantly smaller, MILP problem.   
Let $\mathcal{Q}^*_{rep}$ be the set of repaired queries from the  initial MILP formulation with tuple slicing;
$\mathcal{NC}$ be the set of non-complaint tuples now matching the repaired \texttt{WHERE} clauses, as in Figure~\ref{fig:groups}(b); and $\mathcal{C}^+ = \mathcal{C} \cup \mathcal{NC}$.
We create a new MILP using $\mathcal{C}^+$ as the complaint set.  
The key is to only parameterize the repaired clauses from Step 1 as constraints with undetermined variables.
The variables for all other tuples and queries are fixed to their assigned values from Step 1.
This \emph{refines} the solutions from the previous step while incorporating knowledge about complaints in $\mathcal{NC}$, 
Finally, we use a new objective function to minimize the number of non-complaint tuples 
$t \in \mathcal{NC}$ that are matched by the solution.

In our experiments, we find that this second MILP iteration adds
minimal overhead ($0.1-0.5\%$) with respect to the initial MILP
problem.  In summary, \emph{tuple-slicing} is an effective method to improve the performance of the \emph{basic} approach, without compromising, and often improving, repair quality.

\if{0}
  \subsubsection{A Naive but Flawed approach}
  \ewu{Better explained as: CPLEX searches through an exponential space of all possible combinations of MILP variables.  In a chunked approach, the solution of each chunk is one out of a potentially arbitrary number of possible solutions, thus it is easy to pick an incorrect one}
  A natural idea to optimize the \emph{basic} approach is 
  to \textbf{chunk the query log} into
  smaller, fixed size pieces and then solve each piece at a time: starting
  from the most recent piece, the system linearizes and parameterizes queries 
  in the current piece and derives a corresponding log repair; 
  it then examines the other pieces iteratively
  in the same way. Since complaints only provides
  true values for the most recent database state, in order to avoid 
  linearizing additional queries, 
  we need to know \textbf{rollback} the true values of tuples 
  until the last query in each query log piece. \\
  However, rollback the database is non-easy. An ideal, precise rollback
  algorithm would generate a set of valid ranges for each attribute of a tuple. 
  But the size of valid ranges also grows exponentially with the number queries
  we want to rollback, which, in turn, could not improve the system performance. 
  On the other hand, an approximate, imprecise 
  rollback algorithm would either make the rest of the problems
  infeasible to solve (only maintain fixed number of valid ranges) 
  or result in deriving 
  incorrect log repairs (maintain the lower 
  bound and upper bound among all valid ranges).

  In order to improve the system performance without losing accuracy, we propose
  the following two optimizations: query-slicing optimization 
  based on provenance over queries and
  attribute-slicing optimization based on provenance over 
  attributes. 
\fi

\subsection{Query Slicing: Reducing Queries}
\label{sec:opt:query}

\looseness -1
In practice, many of the queries in the query log could not have affected the \emph{complaint attributes} (defined below). For example, if $q_{N-1}$ and $q_{N}$ 
only read and wrote attribute $A_1$, then they could not have contributed to an error in $A_2$.  
However, if $q_{N}$  wrote $A_2$, then either or both queries may have caused the error. 
In short, if we model a query as a set of attribute read and write operations, those 
not part of the causal read-write chain to the \emph{complaint attributes} can be ignored.
This is the foundation of our \emph{query-slicing} optimization.

\begin{definition} [Complaint Attributes $\mathcal{A}(\mathcal{C})$]
	The set of attributes identified as incorrect in the complaint set.
	\[\mathcal{A}(\mathcal{C}) = \{A_i | t.A_i \neq t^*.A_i, c(t,t^*) \in \mathcal{C}\}\]
\end{definition}

\begin{definition}[Query dependency \& impact]
    Query $q_i$ has \textbf{direct-impact}, $\mathcal{I}(q_i)$, which is
    the set of attributes updated in its modifier function $\mu_{q_i}$
    (e.g., \texttt{SET} clause). Its \textbf{dependency},
    $\mathcal{P}(q_i)$, is the set of attributes involved in its
    condition function $\sigma_{q_i}$.
    We derive the \textbf{full-impact}, $\mathcal{F}(q_i)$, of a query $q_i$ by propagating its direct impact through subsequent queries in the log (Algorithm~\ref{alg:fullimpact}):
    \[
    \mathcal{F}(q_i)=\mathcal{I}(q_i)\bigcup_{\substack{j=i+1\\ \mathcal{F}(q_i)\cap \mathcal{P}(q_j) \neq \emptyset}}^n \mathcal{F}(q_j)
    \]
\end{definition}

By computing the full-impact of $q$, we can determine the extent that it affects $\mathcal{C}$
based on its overlap with the complaint attributes.
Specifically, 
when $|\mathcal{F}(q) \cap \mathcal{A}(C)|=|\mathcal{A}(C)|$, $q$ may affect all complaint attributes and is a candidate for repair; 
when $0 < |\mathcal{F}(q) \cap \mathcal{A}(C)| < |\mathcal{A}(C)|$, 
$q$ contributed to a subset of the complaint attributes and is a candidate for repair;
when $|\mathcal{F}(q) \cap \mathcal{A}(C)|=0$, $q$ is irrelevant 
and can be ignored during the repair process.
We distinguish between the first and second conditions in the special case where we are repairing a \emph{single} 
corrupted query in the query log.  In this case, only queries in the first conditions are candidates for repair because 
the single query must have caused errors in all of the complaint attributes.  This enables \sys to scale significantly better
for this important problem setting. 
Finally, we use $Rel\mathcal{(Q)}$ to denote the set of relevant
queries that are candidates for repair. Our \emph{query slicing}
optimization linearizes only the queries in
$Rel\mathcal{(Q)}$, rather than the entire log, resulting in
smaller problems than the \emph{basic} approach without any loss of accuracy.

\subsection{Attribute Slicing: Reducing Attributes}

In addition to removing irrelevant queries, we additionally avoid encoding irrelevant attributes.
Given $Rel\mathcal{(Q)}$, the relevant attributes can be defined as:
$Rel\mathcal{(A)} = \cup_{q_i \in Rel\mathcal{Q}} (\mathcal{F}(q_i)\cup \mathcal{P}(q_i))$
We propose \emph{attribute slicing} optimization that only encodes constraints for attributes in $Rel\mathcal{(A)}$.
We find that this type of slicing can be effective for wide tables along with queries that focus on a small subset of attributes.

\subsection{Incremental Repairs}\label{sec:incremental}

\looseness -1
Even with the slicing optimizations, the number of undetermined variables can remain high, resulting in slow solver runtime.  
The red bars in Figure~\ref{fig:querysize_vs_time} showed the exponential cost of parameterizing the entire query log as compared to only solving for a single query (blue bars).
These results suggest that it is \emph{faster} to run many small MILP problems than a single large one, and motivates our incremental algorithm.

 \looseness -1
Our \emph{$Inc_k$} approach (Algorithm~\ref{alg:incalg}) focuses on the case where there is a single corrupted query to repair.
It does so by linearizing the full query log, including any slicing optimizations, but only parameterizing and repairing a batch of $k$ consecutive queries at a time. 
This procedure first attempts to repair the $k$ most recent queries, and continues to the next $k$ queries if a repair was not generated.
The algorithm internally calls a modified version of the \emph{basic} approach that takes extra parameters $\{q_i, q_{i+k}\}$, only parameterizes those queries, and fixes the values of all other variables.

The incremental approach prioritizes repairs for complaints that are due to more recent corruptions.
Given that the \emph{basic} algorithm simply fails beyond a small log size, we believe this is a natural and pragmatic assumption to use, and results in 
a $10\times$ scalability improvement.
Our experiments further evaluate different batching level $k$ in the incremental algorithm and show that it is impractical from both a performance and accuracy to have $k > 1$.

\begin{algorithm}[t]
\scriptsize
\caption{$FullImpact:$ Algorithm for finding $\mathcal{F}(q)$.}
\label{alg:fullimpact}
\begin{algorithmic}[2]
\REQUIRE {$\mathcal{Q}$, $q_i$}
\STATE $\mathcal{F}(q_i) \leftarrow \mathcal{I}(q_i)$
\FOR {each $q_j$ in $q_{i+1}, ..., q_{n} \in \mathcal{Q}$}
\IF {$\mathcal{F}(q_i)\cap \mathcal{P}(q_j) \neq \emptyset$}
\STATE $\mathcal{F}(q_i) \leftarrow \mathcal{F}(q_i) \cup \mathcal{F}(q_j)$
\ENDIF
\ENDFOR
\STATE Return $\mathcal{F}(q_i)$
\end{algorithmic}
\end{algorithm}

\begin{algorithm}[t]
\caption{$Inc_k:$ The incremental algorithm. 
}
\scriptsize
\label{alg:incalg}
\begin{algorithmic}[2]
\REQUIRE {$Q, \mathcal{D}_j, \mathcal{D}_n, \mathcal{C}, k$}
\STATE Sort $Q$ from most to least recent
\FOR {each $q_i...q_{i+k} \in Q$}
  \STATE $\mathcal{Q}_{suffix}$ = $\{q_j | j \ge i \}$ 
  \STATE $\mathcal{Q}^*$ $\leftarrow$ $Basic_{params}(\mathcal{Q}_{suffix}, \mathcal{D}_j, \mathcal{D}_n, \mathcal{C}, \{q_i, q_{i+k}\})$
  \IF {$\mathcal{Q}^* \neq \emptyset$}
    \STATE Return $\mathcal{Q}^*$
  \ENDIF
\ENDFOR
\end{algorithmic}
\end{algorithm}

\section{Noisy Complaint Sets}
\label{sec:noise}

As described in the problem setup (Section~\ref{sec:model}), complaint
sets may be imperfect. First, complaint sets are typically incomplete,
missing errors that occur in $D_n$, but are not reported. In this
case, the naive encoding of the query log and database (\naive) will
likely produce an infeasible MILP. In the running example of
Figure~\ref{fig:example}, if the complaint set is incomplete and only
contains a complaint on $t_4$, \naive will interpret $t_3$ as a
correct state and repairing the condition of $q_1$ to a value greater
than $\$86500$ will appear to introduce a new error. The solver will
declare the problem infeasible and will not return a solution.

However, the tuple slicing optimization (Section~\ref{sec:opt:tbsize})
implicitly corrects this problem: By only encoding the tuples in the
incomplete complaint set, the encoded problem does not enforce
constraints on the query's effect on other tuples in the database.
This allows the result to generalize to tuples not in the complaint
set. The second iteration of the MILP execution then uses a soft
constraint on the number of non-complaint tuples that are affected by
the repair in order to address the possibility of over-generalization.

Another possible inaccuracy in the complaint set is the presence of
false positives: some complaints may be incorrectly reporting errors,
or the target tuple $t^*$ of a complaint may be incorrect. This type
of noise in the complaint set can also lead to infeasibility. One can
remove such erroneous complaints as a pre-processing step, using one
of numerous outlier detection algorithms. While this is an
interesting problem, it is orthogonal to the query repair problem that
we are investigating in this work. Thus, in our experiments, we focus
on incomplete complaints sets and assume that there are not erroneous
complaints.

\if{0}
  \subsubsection{False Negatives}
  False negatives are cases when we don't have the full complaint sets, but
  what's provided in the complaint sets are guaranteed as correct. The 
  two-iteration approach in Section~\ref{sec:opt:tbsize} can handle 
  such cases. Refer to Section~\ref{sec:opt:tbsize} for detail.

  \subsubsection{False Positives}
  False positives are cases when we have incorrect information in the complaint sets 
  : they can be a falsely reported tuple which are actually 
  correct, or incorrect suggestions for the erroneous tuples. 
  Including such false positives in the MILP problem may result in a problem
  that is infeasible to solve. Thus, we want to detect and prune these false positives 
  beforehand. However, false positives are hard to 
  detect when there is a lack of information. 
  For example, if we only have few tuples in the complaint
  set, we cannot make any claim about which of them is a false positive 
  complaint. Thus, in this paper, we only considering false positive
  case when the ratio of the number of false positives to the number
  of correctly reported complaints is small.
  A obvious trend for false positives is that they normally are very 
  different or conflict
  with other complaints (Example~\ref{ex:false_positive_1}).
  \begin{example} \label{ex:false_positive_1}
  In Example~r\ref{ex:taxes2}, say there is 
  a false positive complaint on tuple $t_5$ which suggests the correct value
  for $t_5$ at database state $D_3$ should be 
  \{\textbf{ID}:$t_5$, \textbf{rate}: \color{red}{30}
  \color{black}{, \textbf{income}: \$5000, \textbf{owned}
  : }\color{red}{\$1500}\color{black}{\}}. To resolve this complaint, we have
  to guarantee the lower bound of the income range in $q_1$ as 5000. 
  However, the other complaints, $t_1, t_2, t_3$, suggest to
  move this lower bound to at least 9500.0001. In this case, the complaint
  $t_5$ conflicts with the all the other complaints and we may thus
  claim that $t_5$ is likely to be a false positive complaint. 
  \end{example}
  In this section, we introduce a \textbf{Pre-Processing} process that detects 
  false positives effectively. This pre-processing process first searches the 
  best log repair for each complaint separately, it then constructs 
  a bipartite graph between complaints and their impacted tuples and
  searches for the densest sub-graph of the bipartite graph, and finally 
  prunes complaints that are not in the densest sub-graph.
  \begin{itemize}
  \item Using algorithm described in Section~\ref{sec:opt} to solve each 
  complaint individually. Denote the log repair for complaint $c_i$ 
  as $\mathcal{Q}^*(c_i)$, and the impacted tuples of this log repair as
  $T_{c_i}$.
  \item Construct a bipartite graph $G = (\mathcal{C}, T, E)$, where 
  $T = \cup_{i} T_{c_i}$. Note that tuple with same primary key but modified 
  differently are treated as two separate vertices in the bipartite graph. 
  \item Search for densest sub-graph $G' = (\mathcal{C}', T', E')$ in $G$ 
  [] \xlw{cite some papers} and prune complaints in $\mathcal{C} - \mathcal{C}'$. 
  Density[] of a graph $G' = (\mathcal{C}', T', E')$
  is defined by $density(G') = \frac{|E'|}{|\mathcal{C}'|+|T'|}$. 
  \end{itemize}
  For achieve better performance, we can sample tuples
  from the database uniformly when constructing the bipartite graph $G$. 
  We demonstrate how to use this \textbf{Pre-Processing} 
  approach to prune false positive complaint(s) in Example~\ref{ex:false_positive_2}. 
  \begin{example}
  \label{ex:false_positive_2}
  Let's construct the bipartite graph between complaints $t_1, t_2, t_3, t_5$ 
  and tuples in the table for Example~\ref{ex:false_positive_1}. 
  $t_1$ suggests to move the range of income in $q_1$ 
  as $(9000, 10000]$. Similarly, $t_2, t_3, t_5$ suggest $[8570, 90000]$, 
  $[8570, 86000]$, and $[500, 10000]$ respectively. Let's assume tuples are
  uniformly distributed in these ranges. The bipartite graph is shown in 
  Figure~\ref{f:fp}. The density for $t_2, t_3$ is 1.9202 ($\frac{313}{163}$);
  density for $t_2, t_3, t_1$ is 1.9207 ($\frac{315}{164}$); density for 
  $t_1, t_2, t_3, t_5$ is 1.8232 ($\frac{330}{181}$). Thus, $t_5$ is pruned. 
  \begin{figure}[ht]
  \centering
  \includegraphics[width = 0.85\columnwidth]{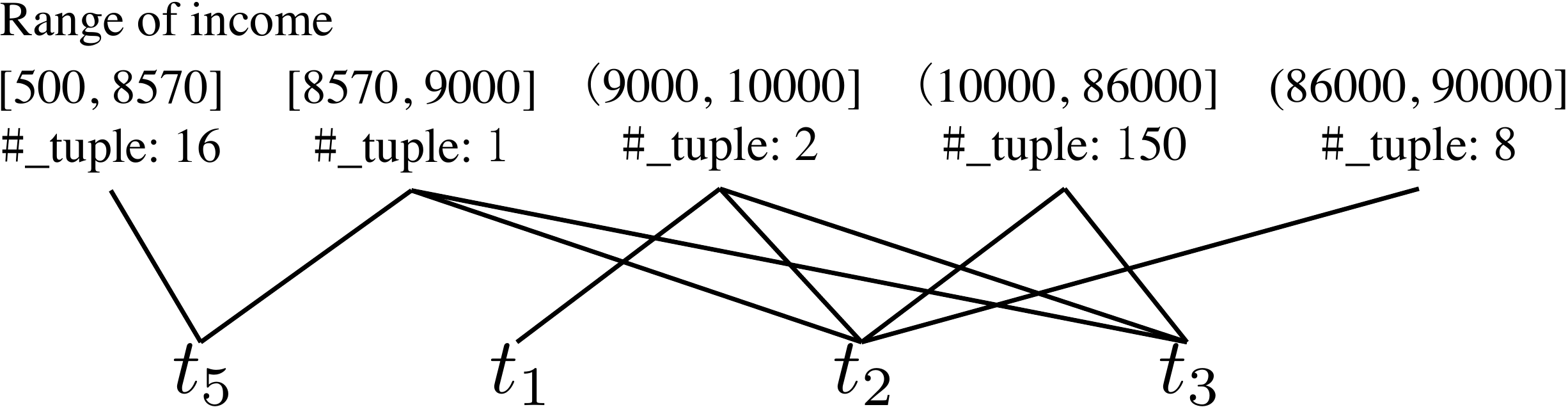}
  \caption{Bipartite graph of complaints and tuples for Example~\ref{ex:taxes2}. }
  \label{f:pf} 
  \end{figure}
  \end{example}

\fi

\section{Experiments}
\label{sec:experiments}

  \begin{figure*}[!htb]
  \hspace*{-.1in}
  \centering
    \vspace*{-.2in}
    \begin{subfigure}[t]{.33\textwidth}
    \includegraphics[width = .99\columnwidth]{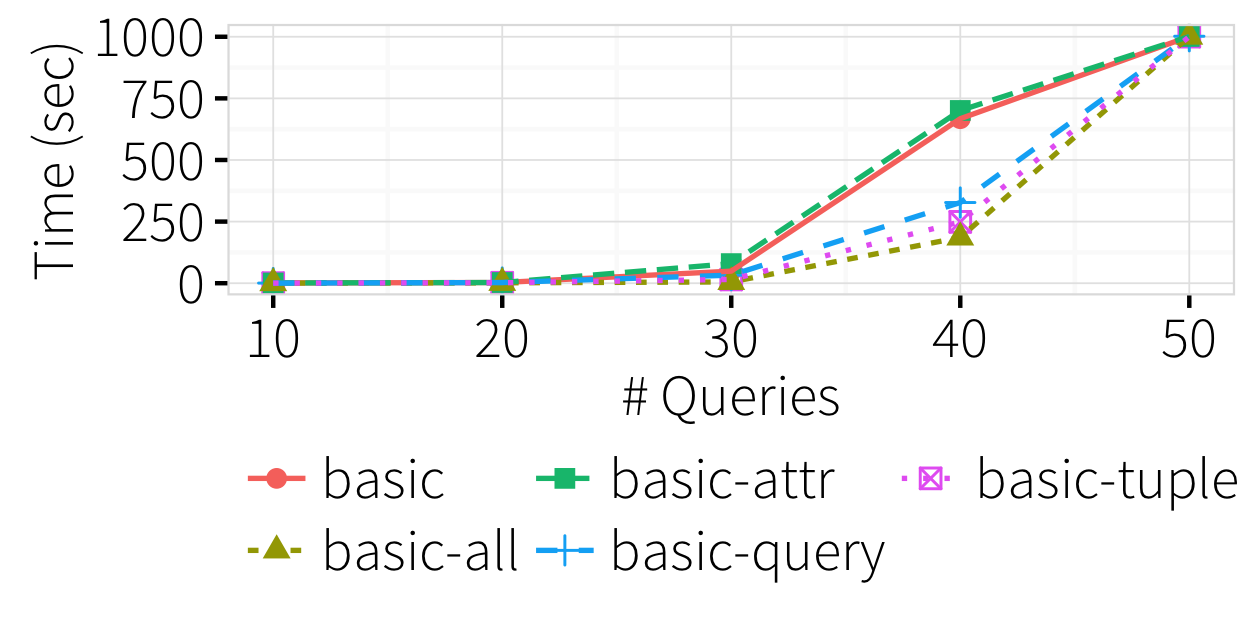}
    \vspace*{-.25in}
    \caption{Performance for multiple corruptions.}
    \label{f:multi_time} 
    \end{subfigure}
    \begin{subfigure}[t]{.33\textwidth}
    \includegraphics[width = .99\columnwidth]{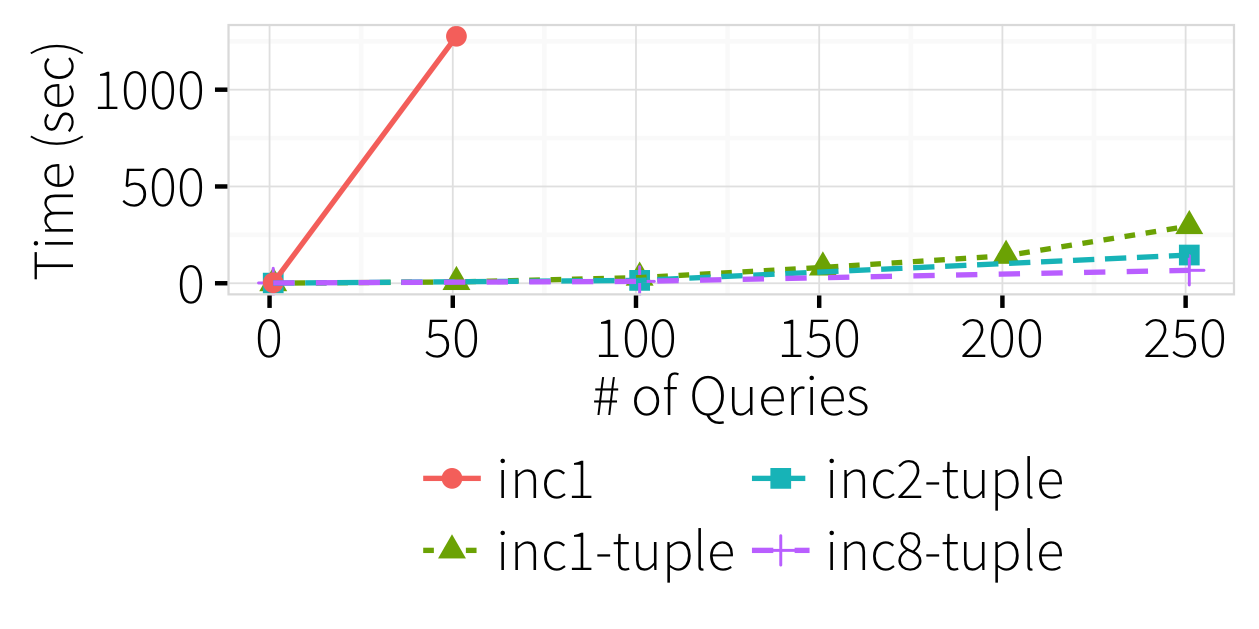}
    \vspace*{-.25in}
    \caption{Performance for single corruption.}
    \label{f:singlequeryinc_time} 
    \end{subfigure}
    \begin{subfigure}[t]{.33\textwidth}
    \includegraphics[width = .99\columnwidth]{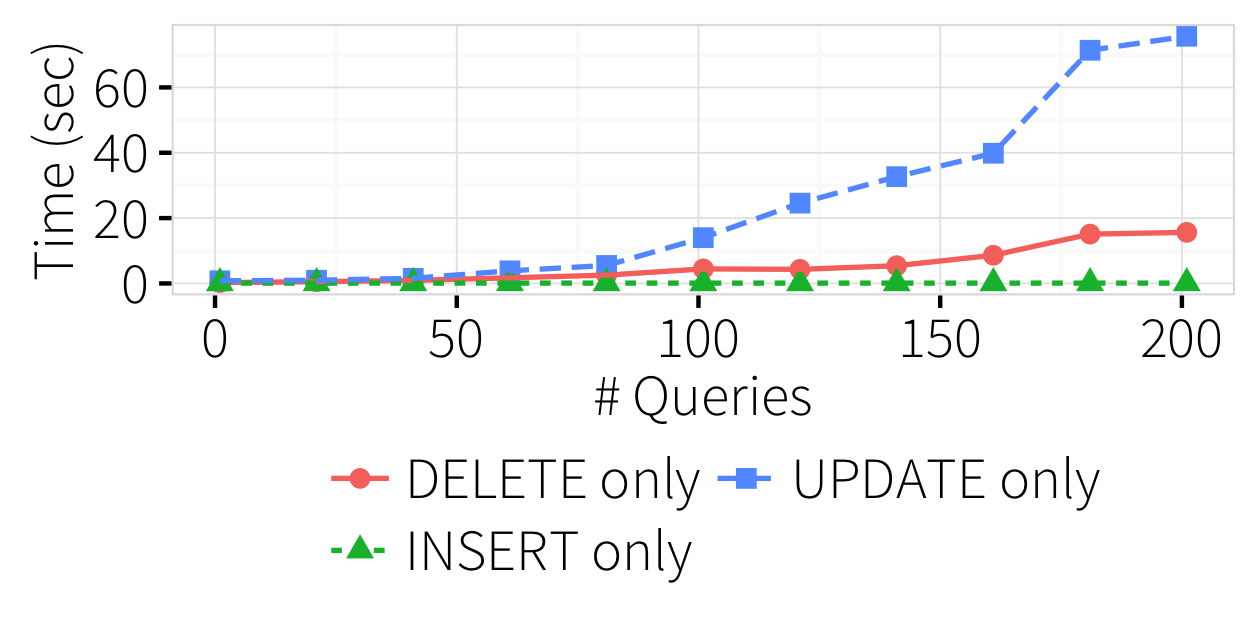}
    \vspace*{-.25in}
    \caption{Performance for different query types.}
    \label{f:indelup_time} 
    \end{subfigure}
    \vspace*{0.2in}
    \\
    \hspace*{-.1in}
    \begin{subfigure}[t]{.33\textwidth}
    \includegraphics[width = .99\columnwidth]{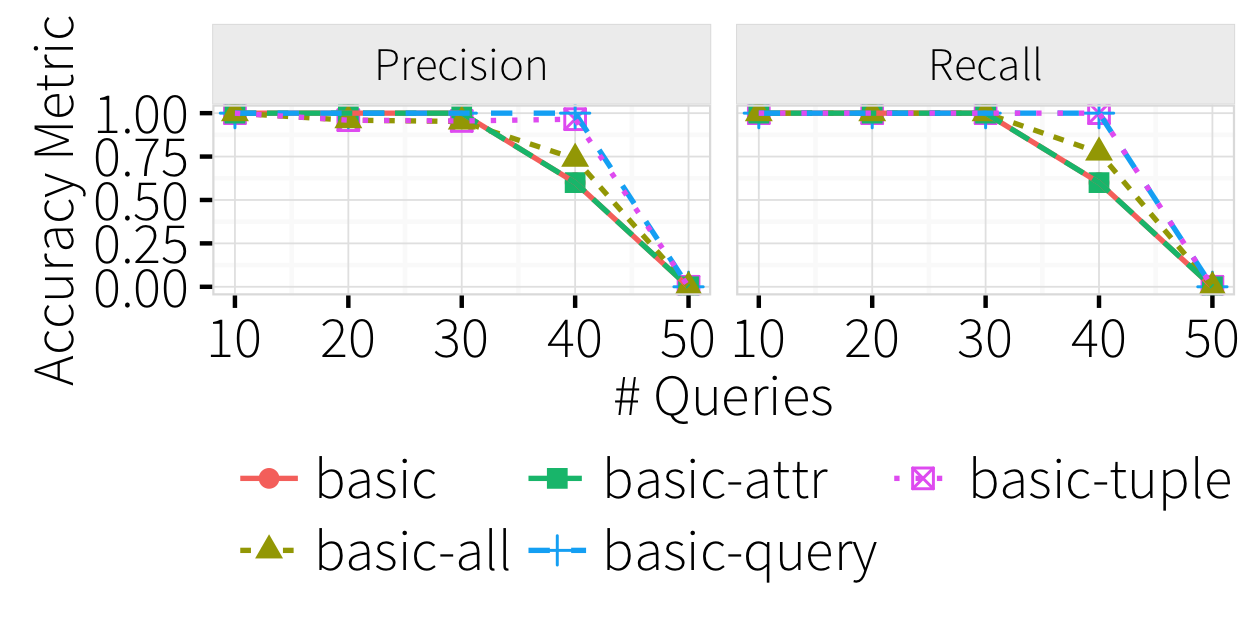}
    \vspace*{-.25in}
    \caption{Accuracy for multiple corruptions.}
    \label{f:multi_acc} 
    \end{subfigure}
    \begin{subfigure}[t]{.33\textwidth}
    \includegraphics[width = .99\columnwidth]{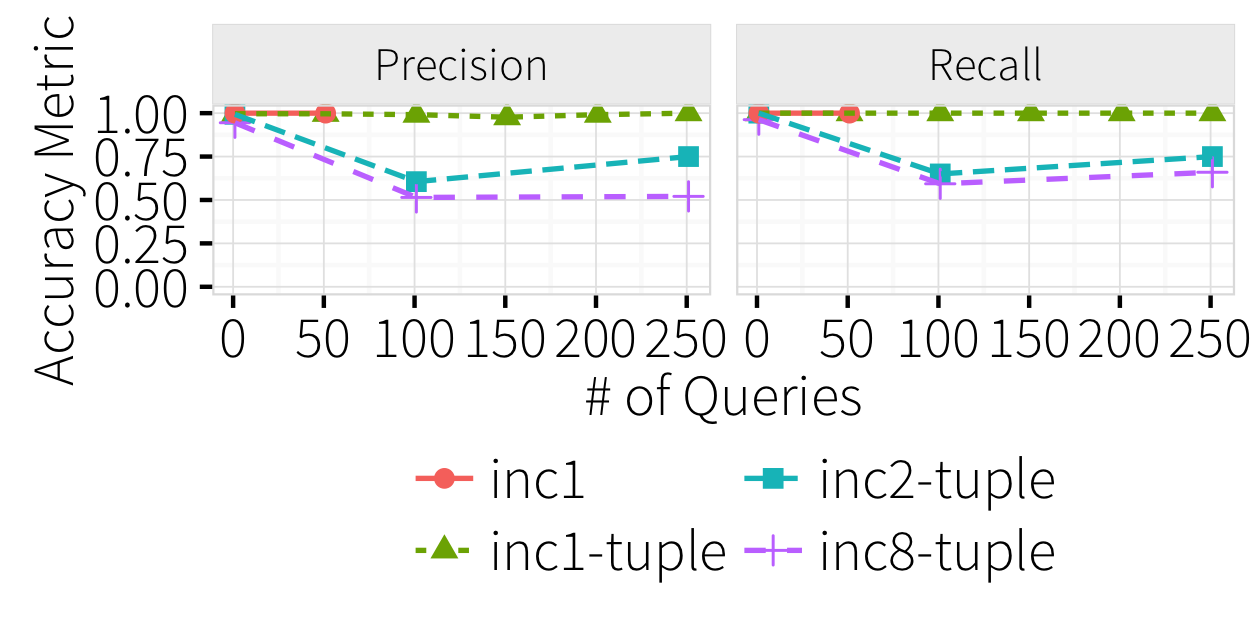}
    \vspace*{-.25in}
    \caption{Accuracy for single corruption.}
    \label{f:singlequeryinc_acc} 
    \end{subfigure}
    \begin{subfigure}[t]{.33\textwidth}
    \includegraphics[width = .99\columnwidth]{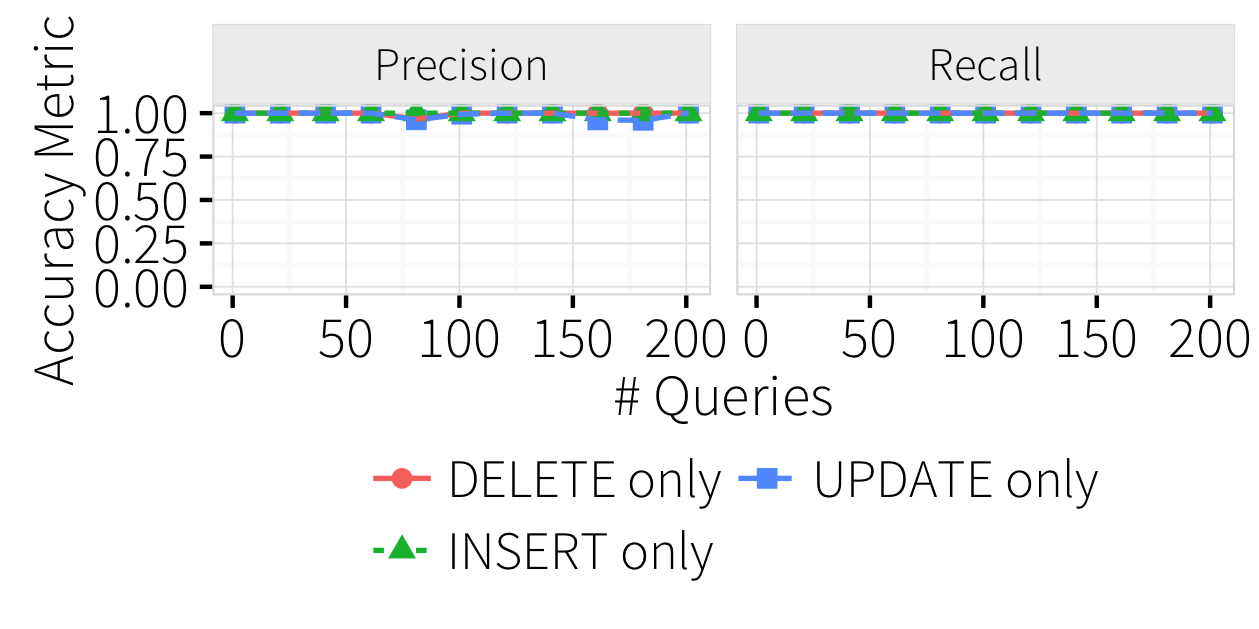}
    \vspace*{-.25in}
    \caption{Accuracy for different query types.}
    \label{f:indelup_acc} 
    \end{subfigure}
    \vspace*{-.1in}
    \caption{Our analysis highlights limitations of \naive, the value of tuple-slicing, and the high cost of \texttt{UPDATE} queries.}
  \end{figure*}

In this section, we carefully study the performance and accuracy
characteristics of the basic MILP-based repair algorithm, 
slicing-based optimizations that improve the latency of the system, 
and the incremental algorithm for single query corruptions. 
Due to the difficulty of collecting corrupt query logs from active deployments, our goal instead is to understand these trade-offs in
controlled synthetic scenarios, as well as study the effectiveness
in typical database query workloads based on widely used benchmarks.

\looseness -1
To this end, our experiments are organized as follows: First, 
we compare the basic and incremental MILP algorithm against the different optimizations 
to highlight the value of different optimizations and the limitations of the basic approach.  
We then show that \texttt{UPDATE} queries are particularly difficult to repair and focus solely
on different types of \texttt{UPDATE}-only workloads to understand how \sys responds to different parameter settings.
We end with an evaluation using established database transaction benchmarks from OLTP-bench~\cite{difallah2013oltp}:
TPC-C~\cite{tpcc} and TATP~\cite{tatp}. 
All experiments were run on 12x2.66 GHz  machines with 16GB RAM running IBM CPLEX~\cite{cplex2014v12} as the MILP solver on CentOS release 6.6.

\subsection{Experimental Setup}

For each of our experiments we generate and corrupt a query log. 
We execute the original and corrupt query logs on an initial (possibly empty) database,
perform a tuple-wise comparison between the resulting database states to generate a true complaint set,
and simulate incomplete complaint sets by removing a subset of the true complaints.
Finally, we execute the algorithms and compare the repaired query log with the true query log, as well as the repaired and true
final database states, to measure performance and accuracy metrics.
Performance is measured as wall clock
time between submitting a complaint set and the system terminating after retrieving a valid repair.  
Accuracy is measured as the repair's precision (percentage of repaired tuples that were correctly fixed), 
the recall (the percentage of the full complaint set that was repaired), 
and the F1 measure (the harmonic mean of precision and recall).
We report the average across 20 runs.
We describe the experimental parameters in the context of the datasets and workloads below.

\stitle{Synthetic:} \label{sec:syntheticgen}
We generate an initial database of $N_D$ random tuples.  
The schema contains a primary key $id$ along with $N_a$ attributes $a_1\ldots a_{N_a}$, whose values are integers picked from $[0, V_d]$ uniformly at random.
We then generate a sequence of $N_q$ queries.
The default setting for these parameters are: $N_D = 1000, N_a = 10, V_d = 200, N_q = 300$.

\texttt{UPDATE} queries are defined by a SET clause that assigns an attribute a $Constant$ or $Relative$ value,
and a WHERE clause can either be a $Point$ predicate on a key, or a $Range$ predicate on non-key attributes:

{\scriptsize
\begin{verbatim}
 SET Clause:                WHERE Clause:
  Constant: SET (a_i=?), ..   Point: WHERE a_j=? & ..
  Relative: SET (a_i=a_i+?)   Range: WHERE a_j in [?,?+r] & ..\end{verbatim}}
\noindent where \verb|?|$\in [0, V_d]$ is random and \verb|r| is the size of the range predicate. 
Query selectivity is by default $2\%$ (\verb|r|$=4$).
Note that a range predicate where \texttt{r = 0} is distinct from a $Point$ predicate due to the non-key attribute.
The WHERE clauses in \texttt{DELETE} queries are generated in an identical fashion, while
\texttt{INSERT} queries insert values picked uniformly at random from $V_d$.
By default, we generate \texttt{UPDATE} queries with non-key range predicates and constant set clauses.
  
In addition, the skew parameter $s$ determines the distribution attributes referenced in the \texttt{WHERE} and \texttt{SET} clauses.  
Each attribute in a query is picked from either a uniform distribution when $s=0$ or a zipfian distribution with exponent $s$.
This allows our experiments to vary between a uniform distribution, where each attribute is
equally likely to be picked, and a skewed distribution where nearly all attributes are the same. 

\looseness -1
\stitle{Benchmarks: } We use the TPC-C~\cite{tpcc} and TATP~\cite{tatp} benchmarks.
The former generates the {\it ORDER} table at scale 1 with one warehouse, and uses the queries that modify the {\it ORDER} table. 
We execute a log of 2000 queries over an initial table containing 6000 tuples.  
$1837$ queries are \texttt{INSERT}s and the rest are \texttt{UPDATE}s. 
The latter TATP workload simulates the caller location system. 
We generate a database from {\it SUBSCRIBER} table with 5000 tuples and $2000$ \texttt{UPDATE} queries.
Both setups were generated using the OLTP-bench~\cite{difallah2013oltp}. 
We introduce a single corruption, and vary corrupted query's index from the most recent query $q_N$ to $q_{N-1500}$.

\stitle{Corrupting Queries:} We corrupt query $q_i$ by replacing it with a randomly
generated query of the same type based on the procedures described above.
To standardize our procedures, we selected a fixed set of indexes $idx$
that are used in all experiments.

\subsection{Preliminary Analysis}
  
\looseness -1
The following set of experiments are designed to establish the rationale for 
the settings in the subsequent experiments.  
Specifically, we compare different slicing-based optimizations of \naive
as the number of queries increases.  
We then evaluate the scalability of the different slicing-based optimizations of the 
incremental approach in the context of a single
corrupted query. We then establish the difficulty of repairing \texttt{UPDATE} 
workloads as compared to other query types. 

\stitle{Multiple Corrupt Queries:} \looseness -1
In this experiment, we compare the basic approach (\naive) against 
each slicing optimization individually($basic-tuple, basic-attr, basic-query$) and all of them together ($basic-all$).  
We use the default settings with $N_D = 1000$ tuples and a sequence of \texttt{UPDATE} queries.
We generate query logs in 5 different sizes $N_q\in \{10, 20, 30, 40, 50\}$ and corrupt 
every tenth query starting from oldest query $q_1$,
up to $q_{41}$.  For example, when the $N_q = {30}$, we corrupt 3 queries: $q_{1,11,21}$. 
We find that the number of queries greatly affects both the scalabality (Figure~\ref{f:multi_time}) 
and the accuracy (Figure~\ref{f:multi_acc}) of the algorithms. Specifically, as the number increases,
the number of possible assignments of the MILP parameters increases exponentially and the solver often takes
longer than our experimental time limit of $1000$ seconds and returns an infeasibility error.  
This is a predominant reason why the accuracy degrades past $30$ queries.  For example, 
when $40$ queries are involved (with $4$ corruptions) 
and we ignore the infeasible executions, the average execution time is $300$ seconds
and the precision and recall are greater than $0.94$.  Unfortunately, with $50$ queries ($5$ corruptions),
all runs exceed the time limit and return infeasibility.

\stitle{Single Corrupt Query:}
In this experiment, we evaluate the efficacy \sys with tuple slicing and incremental optimization
in the special case when one query has been corrupted in a much larger query log. 
We compare \incremental without tuple slicing ($inc_1$) against tuple slicing at 
different batching levels of 1, 2, 8 ($inc_1-tuple; inc_2-tuple, inc_8-tuple$). 
Recall from Section~\ref{sec:incremental} that $inc_k$ parameterizes $k$ consecutive queries in each batch until a repair is found.
Figure~\ref{f:singlequeryinc_time} highlights the scalability limitation of the incremental 
algorithm without tuple-slicing: with 50 queries $inc_1$  easily exceeds the 1000s limit.   
The tuple-slicing scales significantly better (nearly $200\times$ faster), however
the accuracy severely degrades when $k>1$.  
The primary reason is because of infeasibility errors---the MILP problem is much harder, and fails to find a repair.  
This is highlighted by the symmetry between the precision and recall curves.  
A secondary reason is beause the refinement step of  tuple slicing may not generate a fully correct repair and generalize incorrectly.
This is why the precision curve is lower than the recall curve for $inc_8-tuple$.

\stitle{Query Type:}\label{sec:indelup}
Our final preliminary experiment evaluates the incremental algorithm with tuple slicing optimization 
($inc_1-tuple$) on \texttt{INSERT}, \texttt{DELETE}, or \texttt{UPDATE}-only workloads.
We increase the number of queries from $1$ to $200$ and corrupt the oldest query in the log.  
Figure~\ref{f:indelup_time} shows that while the cost of repairing \texttt{INSERT} workloads
remains relatively constant, the cost for \texttt{DELETE}-only and \texttt{UPDATE}-only workloads increase as 
the corruption happens earlier in the query log---and a much faster rate for \texttt{UPDATE} queries.
The F1 score for all settings is nearly 1 (Figure~\ref{f:indelup_acc}).
 \begin{figure*}[!htb]
    \vspace*{-.1in}
    \centering
    \begin{subfigure}[t]{.49\textwidth}
    \includegraphics[width = .99\columnwidth]{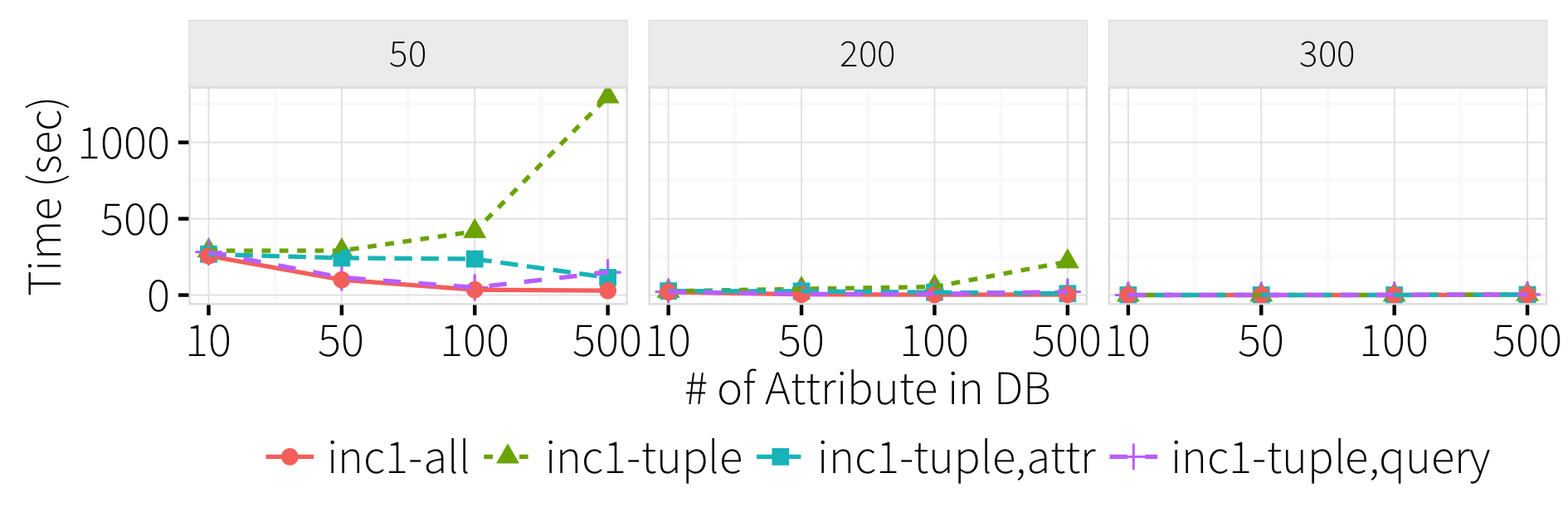}
    \vspace*{-.1in}
    \caption{\# of attributes vs time.}
    \label{f:attr} 
    \end{subfigure}
    \begin{subfigure}[t]{.49\textwidth}
    \includegraphics[width = .99\columnwidth]{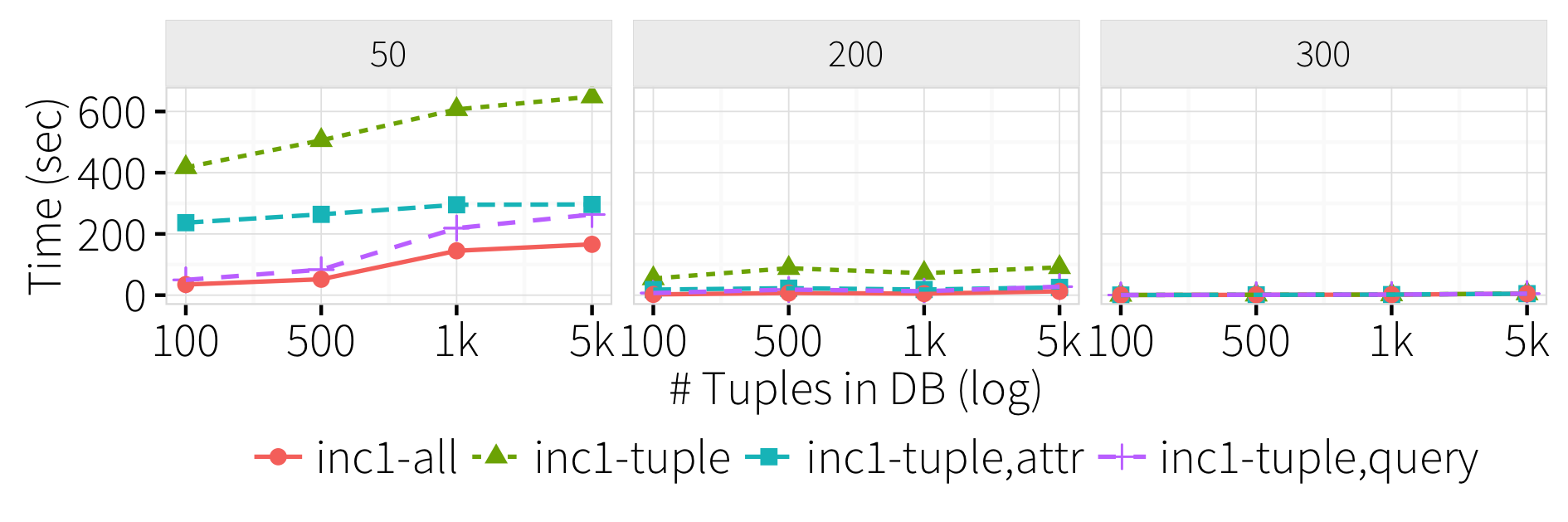}
    \vspace*{-.1in}
    \caption{Database size vs time ($N_a = 100$)}
    \label{f:attr100} 
    \end{subfigure}
    \vspace*{-.1in}
    \caption{For datasets with many attributes, the optimizations result in significant improvements.}
  \end{figure*}
\smallskip
\emph{Takeaways: We find that \naive, even with slicing optimizations,
has severe scalability limitations due to the large number of undetermined values---this is unsurprising as MILP constraint solving is an NP-hard problem.
In contrast, the incremental algorithms can scale to larger query log sizes, however only a batch size of $k=1$ can retain high repair quality.
\texttt{UPDATE} queries translate into more undetermined variables than other query types, and are significantly more expensive to repair.
Based on these results, we focus on the incremental algorithm $inc_1$ and the more difficult \texttt{UPDATE}-only workloads.
}

\subsection{Synthetic Incremental Experiments}
\label{sec:experiments:synth}
We divide these experiments into two groups: the first evaluates the different slicing optimizations under different settings; 
the latter algorithm and varies workload and dataset parameters.
Note that omit accuracy figures when the F1-score is $\ge 0.99$.

\stitle{Comparing optimizations.}

\emph{Varying \# Attributes:} \looseness -1
We first evaluate \sys and its optimizations by increasing the number of attribute ($N_a \in [10, 500]$) with $N_D = 100$.
As shown in Figure~\ref{f:attr}, when the number of attribute in a table is small (e.g., $N_a=10$) all algorithms appear identical. 
However, increasing the number of attribute exhibits a larger benefit for query and attribute slicing (up to $6.8\times$ reduction compared to tuple-slicing).
When the table is wide ($N_a = 500$), applying all optimizations ($inc_1-all$) is $40\times$ faster than \emph{tuple-slicing} alone.  

\emph{Database Size:} 
We vary the database size ($N_D \in [100,5000]$) with a large number of attributes ($N_a = 100$).
We fix the number of complaints by decreasing the query selectivity in proportion to $N_D$'s increase---the specific mechanism to do so does not affect the findings.
Figure~\ref{f:attr100} shows that the costs are relatively flat until the corruption occurs in an old query ($q_{50}$).  
In addition, we find that the cost is highly correlated with the number of candidate queries that are encoded in the MILP problem.
The increase in cost despite tuple-slicing is due to the increasing number of candidate queries in the system; 
we believe this increasing trend is due to the solver's ability to prune constraints that correspond to queries that clearly will not affect the complaint set---an implicit form of query slicing.  
Applying attribute-slicing supercedes this implicit optimization and results in a flat curve, 
while query-slicing explicitly reduces the number of candidate queries in proportion with the database size, and leads to the increasing trend.
Ultimately, combining all three optimizations improves the latency over tuple-slicing by $2\times$ in the worst case, and up to $4\times$ in the best case.

\stitle{Sensitivity to data and workload factors.}

The following set of synthetic experiments focus on a single \sys setting---incremental with tuple-slicing---and
individually vary numerous database and workload parameters in order to tease apart the algorithm performance.  
These include factors such as query complexity, log size $N_q$, database size $N_D$ and the skew of the query predicates.
We focus on a narrow table setting that contains $N_a = 10$ attributes, and a single corrupt query in the query log.

\begin{figure*}[!htb]
    \hspace*{-.1in}
    \centering
     \begin{subfigure}[t]{.33\textwidth}
      \includegraphics[width = .99\columnwidth]{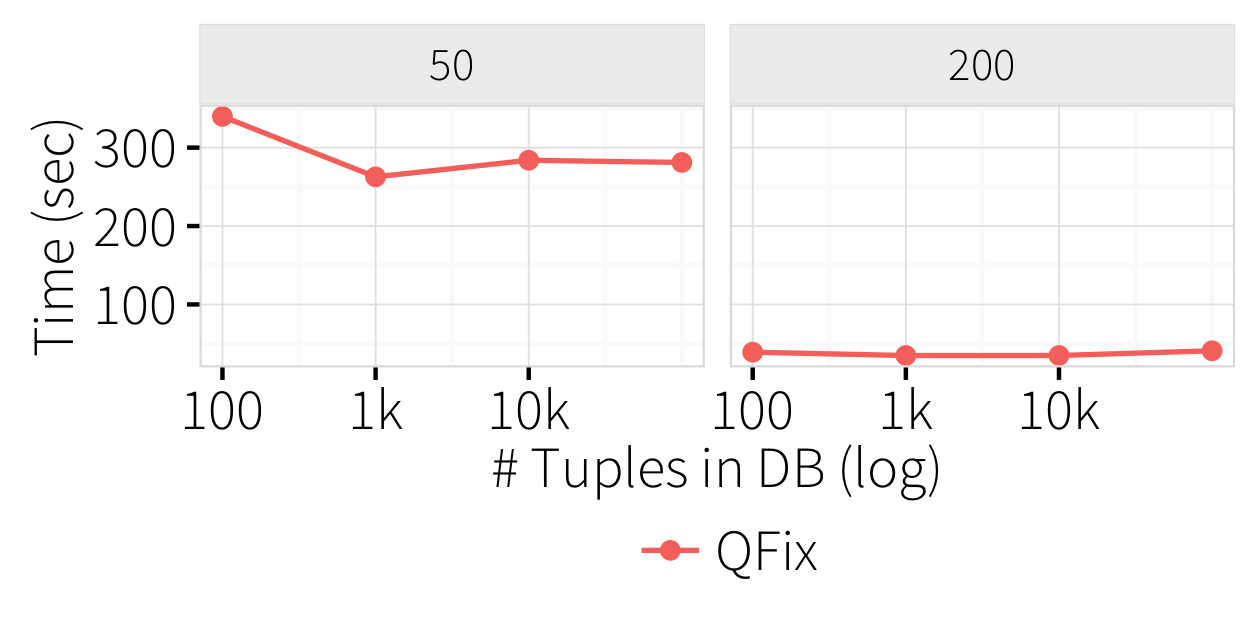}
      \vspace*{-.25in}
      \caption{Database size vs time.}
      \label{f:dbsize_time} 
    \end{subfigure}
    \begin{subfigure}[t]{.33\textwidth}
      \includegraphics[width = .99\columnwidth]{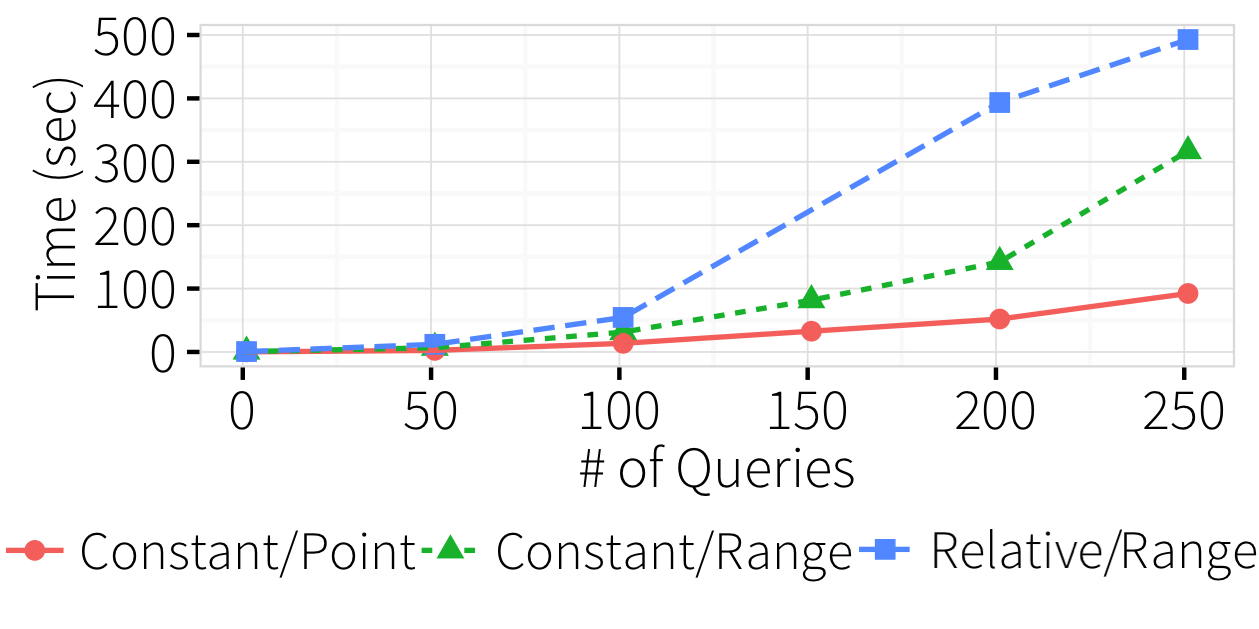}
      \vspace*{-.25in}
      \caption{Performance of diff. query clause types.}
      \label{f:qidx_time} 
    \end{subfigure}
    \begin{subfigure}[t]{.33\textwidth}
      \includegraphics[width = .99\columnwidth]{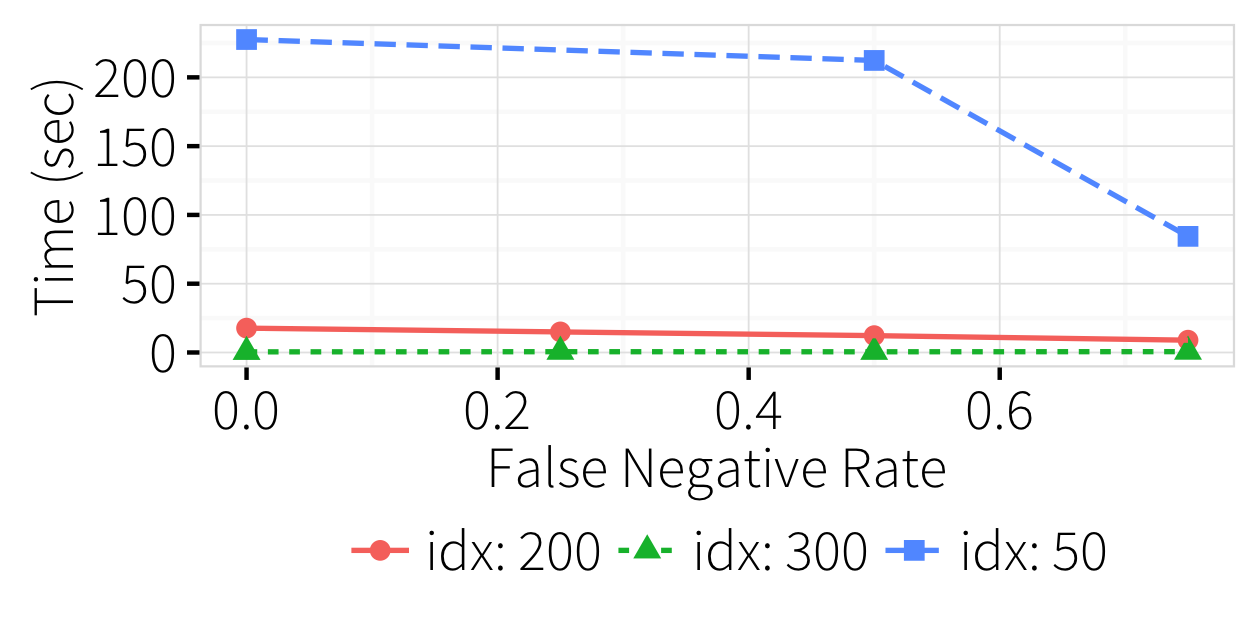}
      \vspace*{-.25in}
      \caption{False negatives vs time.}
      \label{f:falsenegative_time} 
    \end{subfigure} 
    \\
    \hspace*{-.1in}
    \vspace*{.2in}
    \begin{subfigure}[t]{.33\textwidth}
      \includegraphics[width = .99\columnwidth]{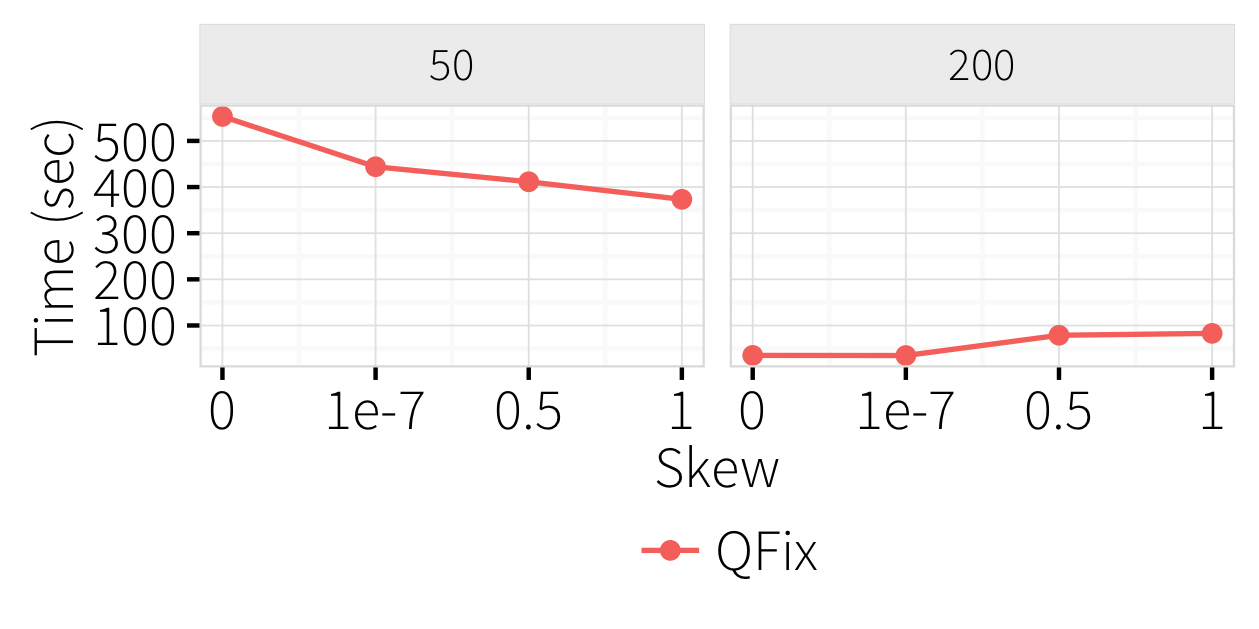}
      \vspace*{-.25in}
      \caption{Skew vs time.}
      \label{f:skew_time} 
    \end{subfigure}
    \begin{subfigure}[t]{.33\textwidth}
      \includegraphics[width = .99\columnwidth]{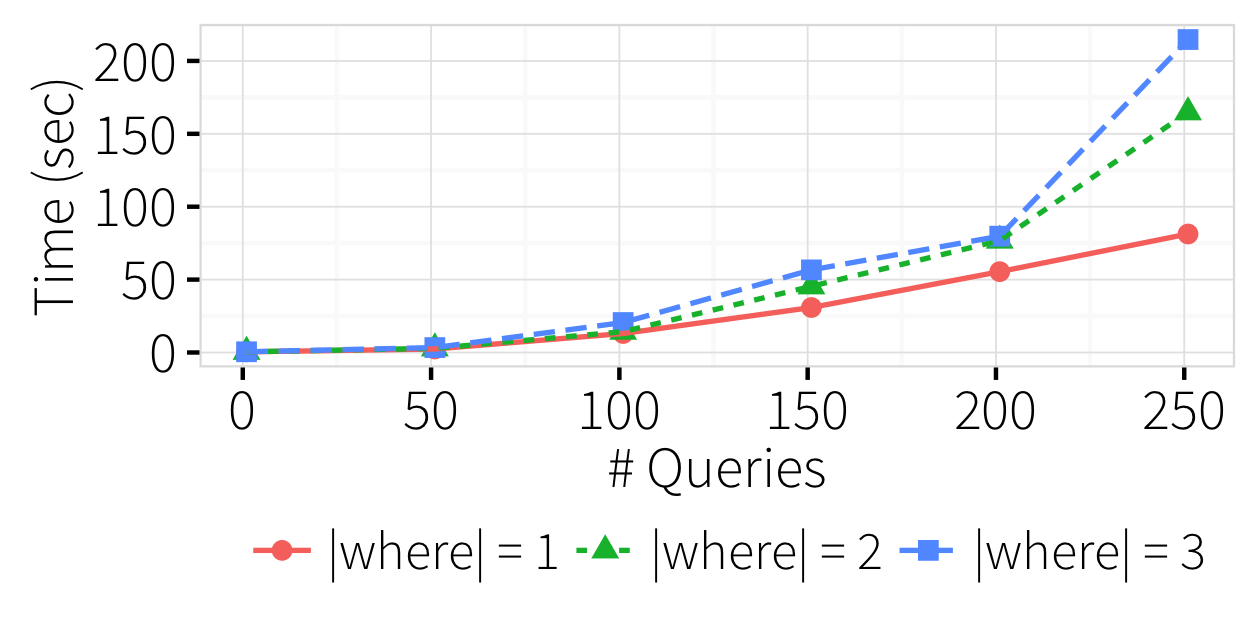}
      \vspace*{-.25in}
      \caption{Query dimensionality vs time.}
      \label{f:where_time} 
    \end{subfigure}
    \begin{subfigure}[t]{.33\textwidth}
      \includegraphics[width = .99\columnwidth]{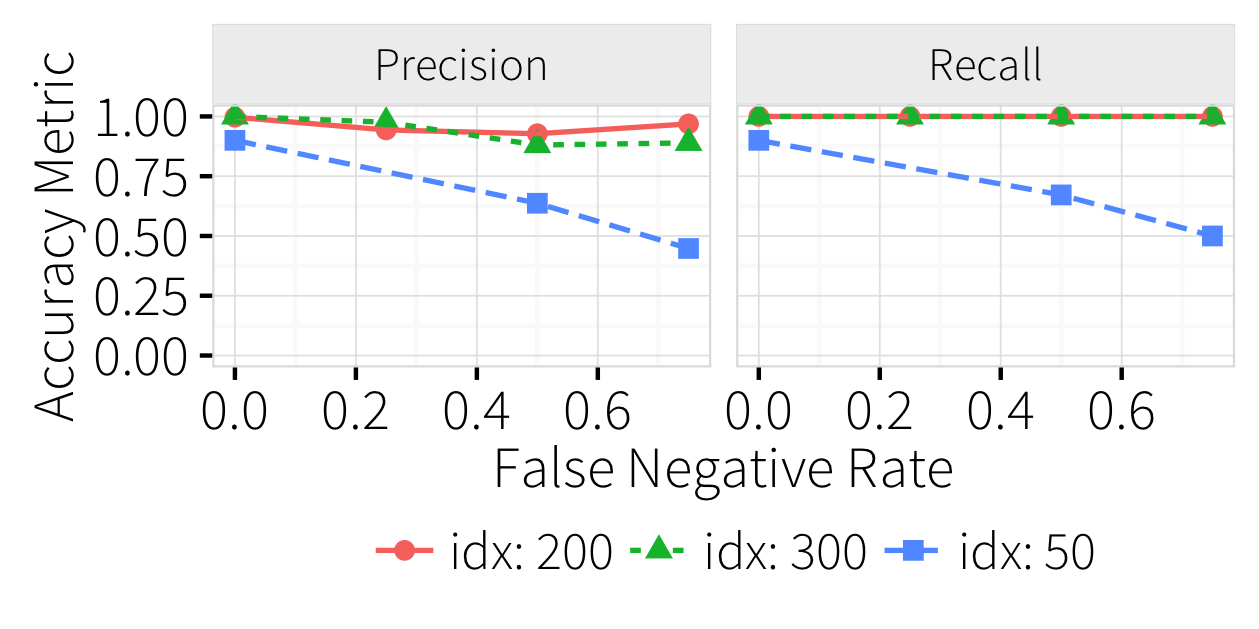}
      \vspace*{-.25in}
      \caption{False negatives vs accuracy.}
      \label{f:falsenegative_acc} 
    \end{subfigure}
    \vspace*{-.25in}
    \caption{
    Performance and accuracy can be sensitive to some data and workload parameters, such as query skew, clause types, and dimensionality.  Notably, \sys can handle high false negative rates when the error resides in recent queries.}
  \end{figure*}

\emph{Database Size:} \looseness -1
Figure~\ref{f:dbsize_time} varies the database size ($N_D \in [100, 100k]$), and fixes query output cardinality and complaint set size in the same way as the previous scalability experiment.
In contrast to the previous experiment, the scalabality curve is nearly flat for both corruption query indices.
The reason is because the solver's implicit pruning optimization is less effective when there are only $10$ attributes: Every query is likely to touch an attribute that affects the complaint set.
We verified this by applyng query-slicing to the same setting, and found far fewer queries were pruned compared to the previous experiment.
It takes less than a minute to perfectly repair the recent corruption $q_{200}$,
and around $4$ minutes for the older corruption $q_{50}$, even as the database size increases.
At smaller database sizes, the randomness in the workload generator leads to variability in the size of the complaint set, and ultimately a larger and more difficult MILP problem.
The exponential relationship between solver time and problem difficulty results in the higher average latency for $q_{50}$.

\emph{Query Clause Type: }
So far, we have focused on \texttt{UPDATE} queries with constant set clauses and range predicates.  
Figure~\ref{f:qidx_time} individually varies each clause and compares against {\it Constant/Point} and {\it Relative/Range} queries. 
The x-axis varies the index of the corrupted query between $q_1$ and $q_{249}$.
We find that point predicates and constant set clauses are easier to solve than range predicates and relative set clauses, respectively.
The reason for the former pair is because range predicates double the number of undetermined variables as compared to point queries.  
In addition, point queries are on key attributes, thus further reduces the possible search space.  
We believe the latter pair is because the constant set clauses break the causal relationship between input and output records for the overwritten values.
This both simplifies the difficulty of the constraint problem, and reduces the total number of constraints.

\emph{Predicate Dimensionality:}
Figure~\ref{f:where_time} varies the dimensionality of the update queries by increasing the number of predicates in the \texttt{WHERE} clause, while keeping the query cardinality constant.
The cost increases with the dimensionality because each additional predicate is translated into a new set of constraints and undetermined variables, increasing the problem complexity.

\emph{Skew:} We now study the effects of attribute skew on the algorithms.
We increase the skew parameter from $0$ (uniform) to $1$ (nearly every attribute is $A_0$) 
and find a reduction in latency (Figure~\ref{f:skew_time}).
We believe the reason is because increasing the skew focuses the query predicates over a smaller set of logical attributes, 
and increases the number of constraints placed on each of the logical attributes used in the query log.  
Each of these constraints reduces the search space of allowable values for that attribute, and thus simplifies the MILP problem.
This result suggests that \sys may be well suited for many transaction systems that naturally exhibit query skew.
Note that the overall number of constraints in the problem is the same, only their distribution over the query attributes has changed.

\emph{Incomplete Complaint Set:}
Our final experiment (Figures~\ref{f:falsenegative_time} and~\ref{f:falsenegative_acc}) varies the fase positive rate in incomplete complaint sets.
We increase the rate from $0$ ($0\%$ missing in the complaint set) to $.75$ ($75\%$ are missing).  
We find that reducing the size of the submitted complaint set naturally improves the repair performance,
however the repair quality, both precision and recall in Figure~\ref{f:falsenegative_acc}, may suffer if the corruption occured in a very old query. 
This is expected because \sys targets reported complaints, thus unreported complaints may easily be missed and lead to low recall.
In addition, despite the refinement step of tuple-slicing, the repair may over generalize and ``fix'' the wrong records, leading to low precision.

\smallskip
\textit{Takeaways: we find that the performance of the different repair algorithm 
heavily depends on the property of the datasets and queries---in particular, the number of encoded candidate queries and the number of attributes. 
Attribute and query slicing show significant gain for datasets with large number of attributes.
\sys is able to solve hard problems (with ~$200$ \texttt{UPDATE}-only queries) in seconds or minutes---particularly if the error is recent. }

\subsection{Benchmarks}
\label{sec:experiments:benchmark}

\begin{figure}[!htb]
\centering
  \includegraphics[width = .75\columnwidth]{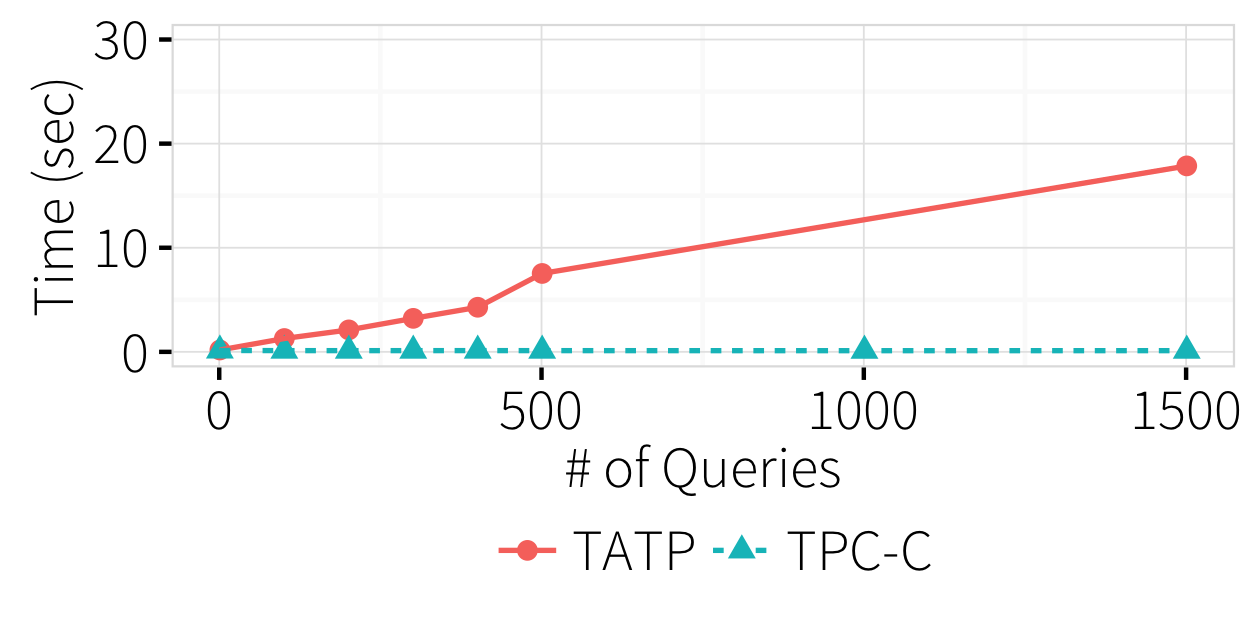}
  \vspace*{-.2in}
  \caption{Our experiments on benchmark OLTP workloads show that \sys can produce repairs very fast.}
  \label{f:tpcctatp} 
\end{figure}

Figure~\ref{f:tpcctatp} plots the performance of the incremental algorithm using tuple-slicing on the TPC-C and TATP benchmark applications.  
The key reason is that each query affects a small set of records and leads to a very small complaint set---$1$ or $2$ on average.
In addition, tuple and query slicing can aggressively reduce the total number of constraints to a very small number---often less than $100$ in total.
Furthermore for TPC-C, the queries are predominantly \texttt{INSERT} queries, which \sys can solve within milliseconds.
Finally, we evaluated \sys on Example~\ref{ex:taxes} in Figure~\ref{fig:example} and fully repaired the correct query in 35 milliseconds.

\smallskip
{\it Takeaways: many workloads in practice are dominated by \texttt{INSERT} and point \texttt{UPDATE} queries (ignoring the dominant percentage of read-only queries).  
In these settings, \sys is very effective at reducing the number of constraints and can derive repairs with near-interactive latencies.}

\section{Related Work}
\label{s:related}
\sys tackles the problem of diagnosis and repair in relational query
histories (query logs). It does not aim to correct errors in the data
directly, but rather to find the underlying reason for reported errors
in the queries that operated on the data. This contrasts with
traditional data 
cleaning~\cite{dallachiesa2013nadeef,Abiteboul99,Koudas2006,Galhardas2000
} which oftentimes focuses on identifying and correcting data
``in-place.'' as well as cleaning techniques that provide repairs for the
identified errors~\cite{Fan2008b, ChuIP13, Beskales2010, Cong2007, Chalamalla2014}.
By analyzing the process that generate the errors,  \sys can help detect and repair systemic 
errors that have not been explicitly identified.

Several tools~\cite{GolabKKS10, wang2015} explore systemic reasons for errors 
and generate feature sets or patterns of attributes that characterize those errors.
However, such techniques are oblivious to the actual executed queries and do not provide fixes. 

The topic of query revisions has been studied in the context of
why-not explanations~\cite{Chapman2009, tran2010conquer,tzompanaki14semi } and explaining query ouputs~\cite{Wu13,GebalyAGKS14,Roy2014}. 
But all these approaches are
limited to selection predicates of \texttt{SELECT} queries, and they
only typically consider one query at a time.

Finally, as \sys traces errors in the queries that manipulate data, it
has connections to the field of \emph{data and workflow provenance}.
our algorithms build on several formalisms introduced by work in this
domain. These formalisms express why a particular data item appears in
a query result, or how that query result was produced in relation to
input data~\cite{BunemanKT01,GKT07-semirings, CheneyCT09, CuiWW00
}.

\section{Summary and discussion}

The general problem of data errors is highly complex, and exacerbated by its highly contextual nature.
We believe that an approach that explains and repairs such data errors based on operations performed by the application or user
is a promising step towards incorporating contextual hints into the analysis process.

Towards this goal, we presented \sys, the first framework to diagnose and
repair errors in the queries that operate on the data.
Datasets are typically dynamic: even if a dataset starts clean,
updates may introduce new errors. \sys can
analyze query logs to trace reported errors to the queries that
introduced them. This in turn helps identify additional errors
in the data that may have been missed and gone unreported.

We proposed a basic algorithm, \texttt{basic}, that uses non-trivial transformation rules to
encode information from the data and the query log as a MILP problem. We further improve 
\texttt{basic} with two types of optimizations: 
(1)~slicing-based optimizations that reduce the problem
size without compromising the accuracy, but rather often improving it, and 
(2)~an incremental approach that analyzes a single query at a time. 
Our experiments show that the latter
optimization can achieve significant scaling gains for single-query
errors, without significant reduction in accuracy.

To the best of our knowledge, \sys is the first formalization and solution to the diagnosis
and repair of errors using past executed queries. 
Obviously, correcting such errors in practice poses additional challenges. 
The initial version of \sys described in this paper focuses on a constrained problem consisting of
simple (no subqueries, UDFs, aggregations, nor joins)
single-query transactions with clauses composed of linear functions, and
complaint sets without false positives.
In future work, we hope to extend our techniques to relax these limitations.
In addition, we plan to investigate additional methods of scaling the constraint analysis, 
as well as techniques that can adapt the benefits of single-query analysis to errors in multiple queries.

\balance

{
\bibliographystyle{abbrv}
\bibliography{main}

\begin{thebibliography}{10}

\bibitem{Abiteboul99}
S.~Abiteboul, S.~Cluet, T.~Milo, P.~Mogilevsky, J.~Sim{\'e}on, and S.~Zohar.
\newblock Tools for data translation and integration.
\newblock In {\em IEEE Data Engineering Bulletin}, 1999.

\bibitem{Amsterdamer2011}
Y.~Amsterdamer, S.~B. Davidson, D.~Deutch, T.~Milo, J.~Stoyanovich, and
  V.~Tannen.
\newblock Putting lipstick on pig: Enabling database-style workflow provenance.
\newblock In {\em PVLDB}, 2011.

\bibitem{Beskales2010}
G.~Beskales, I.~F. Ilyas, and L.~Golab.
\newblock Sampling the repairs of functional dependency violations under hard
  constraints.
\newblock In {\em PVLDB}, 2010.

\bibitem{BunemanKT01}
P.~Buneman, S.~Khanna, and W.~C. Tan.
\newblock Why and where: A characterization of data provenance.
\newblock In {\em ICDT}, 2001.

\bibitem{Chalamalla2014}
A.~Chalamalla, I.~F. Ilyas, M.~Ouzzani, and P.~Papotti.
\newblock Descriptive and prescriptive data cleaning.
\newblock In {\em SIGMOD}, 2014.

\bibitem{Chapman2009}
A.~Chapman and H.~V. Jagadish.
\newblock Why not?
\newblock In {\em SIGMOD}, 2009.

\bibitem{Chen2011}
S.~Chen, X.~L. Dong, L.~V. Lakshmanan, and D.~Srivastava.
\newblock We challenge you to certify your updates.
\newblock In {\em SIGMOD}, 2011.

\bibitem{CheneyCT09}
J.~Cheney, L.~Chiticariu, and W.~C. Tan.
\newblock Provenance in databases: Why, how, and where.
\newblock In {\em Foundations and Trends in Databases}, 2009.

\bibitem{ChuIP13}
X.~Chu, I.~F. Ilyas, and P.~Papotti.
\newblock Holistic data cleaning: Putting violations into context.
\newblock In {\em ICDE}, 2013.

\bibitem{Cong2007}
G.~Cong, W.~Fan, F.~Geerts, X.~Jia, and S.~Ma.
\newblock Improving data quality: Consistency and accuracy.
\newblock In {\em VLDB}, 2007.

\bibitem{tpcc}
T.~T.~P. Council.
\newblock Benchmark c: Standard specification (revision 5.9.0), 2014.

\bibitem{cplex2014v12}
I.~CPLEX.
\newblock High-performance software for mathematical programming and
  optimization, 2005.

\bibitem{CuiWW00}
Y.~Cui, J.~Widom, and J.~L. Wiener.
\newblock Tracing the lineage of view data in a warehousing environment.
\newblock In {\em ACM Transactions on Database Systems}, 2000.

\bibitem{dallachiesa2013nadeef}
M.~Dallachiesa, A.~Ebaid, A.~Eldawy, A.~Elmagarmid, I.~F. Ilyas, M.~Ouzzani,
  and N.~Tang.
\newblock Nadeef: a commodity data cleaning system.
\newblock In {\em SIGMOD}, 2013.

\bibitem{difallah2013oltp}
D.~E. Difallah, A.~Pavlo, C.~Curino, and P.~Cudre-Mauroux.
\newblock Oltp-bench: An extensible testbed for benchmarking relational
  databases.
\newblock In {\em VLDB}, 2013.

\bibitem{oltpbench}
D.~E. Difallah, A.~Pavlo, C.~Curino, and P.~Cudre-Mauroux.
\newblock Oltp-bench: An extensible testbed for benchmarking relational
  databases.
\newblock In {\em VLDB}, 2013.

\bibitem{eckerson2002}
W.~W. Eckerson.
\newblock Data quality and the bottom line.
\newblock {\em TDWI Report, The Data Warehouse Institute}, 2002.

\bibitem{Fan2008}
W.~Fan, F.~Geerts, and X.~Jia.
\newblock A revival of integrity constraints for data cleaning.
\newblock In {\em VLDB}, 2008.

\bibitem{Fan2008b}
W.~Fan, F.~Geerts, X.~Jia, and A.~Kementsietsidis.
\newblock Conditional functional dependencies for capturing data
  inconsistencies.
\newblock In {\em ACM Transactions on Database Systems}, 2008.

\bibitem{galar2012review}
M.~Galar, A.~Fernandez, E.~Barrenechea, H.~Bustince, and F.~Herrera.
\newblock A review on ensembles for the class imbalance problem: bagging-,
  boosting-, and hybrid-based approaches.
\newblock {\em Systems, Man, and Cybernetics, Part C: Applications and Reviews,
  IEEE Transactions on}, 42(4):463--484, 2012.

\bibitem{Galhardas2000}
H.~Galhardas, D.~Florescu, D.~Shasha, and E.~Simon.
\newblock Ajax: An extensible data cleaning tool.
\newblock In {\em SIGMOD}, 2000.

\bibitem{GebalyAGKS14}
K.~E. Gebaly, P.~Agrawal, L.~Golab, F.~Korn, and D.~Srivastava.
\newblock Interpretable and informative explanations of outcomes.
\newblock In {\em PVLDB}, 2014.

\bibitem{GolabKKS10}
L.~Golab, H.~J. Karloff, F.~Korn, and D.~Srivastava.
\newblock Data auditor: Exploring data quality and semantics using pattern
  tableaux.
\newblock In {\em PVLDB}, 2010.

\bibitem{Grady13}
B.~Grady.
\newblock Oakland unified makes \$7.6{M} accounting error in budget; asking
  schools not to count on it.
\newblock In {\em Oakland}, 2013.

\bibitem{GKT07-semirings}
T.~J. Green, G.~Karvounarakis, and V.~Tannen.
\newblock Provenance semirings.
\newblock In {\em PODS}, 2007.

\bibitem{he2009learning}
H.~He and E.~A. Garcia.
\newblock Learning from imbalanced data.
\newblock {\em Knowledge and Data Engineering, IEEE Transactions on},
  21(9):1263--1284, 2009.

\bibitem{Khoussainova2006}
N.~Khoussainova, M.~Balazinska, and D.~Suciu.
\newblock Towards correcting input data errors probabilistically using
  integrity constraints.
\newblock In {\em MobiDE}, 2006.

\bibitem{Koudas2006}
N.~Koudas, S.~Sarawagi, and D.~Srivastava.
\newblock Record linkage: Similarity measures and algorithms.
\newblock In {\em SIGMOD}, 2006.

\bibitem{tiresias}
A.~Meliou and D.~Suciu.
\newblock Tiresias: The database oracle for how-to queries.
\newblock In {\em SIGMOD}, 2012.

\bibitem{quinlan1987}
J.~R. Quinlan.
\newblock Simplifying decision trees.
\newblock {\em International journal of man-machine studies}, 27(3):221--234,
  1987.

\bibitem{Roy2014}
S.~Roy and D.~Suciu.
\newblock A formal approach to finding explanations for database queries.
\newblock In {\em SIGMOD}, 2014.

\bibitem{sakalerrors}
M.~Sakal and L.~Rakovi{\'{c}}.
\newblock Errors in building and using electronic tables: Financial
  consequences and minimisation techniques.
\newblock In {\em Strategic Management}, 2012.

\bibitem{tran2010conquer}
Q.~T. Tran and C.-Y. Chan.
\newblock How to conquer why-not questions.
\newblock In {\em SIGMOD}, 2010.

\bibitem{tzompanaki14semi}
K.~Tzompanaki, N.~Bidoit, and M.~Herschel.
\newblock Semi-automatic sql debugging and fixing to solve the missing-answers
  problem.
\newblock In {\em VLDB PhD Workshop}.

\bibitem{wang2015}
X.~Wang, X.~L. Dong, and A.~Meliou.
\newblock Data x-ray: A diagnostic tool for data errors.
\newblock In {\em SIGMOD}, 2015.

\bibitem{qfixarxiv}
X.~Wang, A.~Meliou, and E.~Wu.
\newblock Qfix: Diagnosing errors through query histories.
\newblock {\em CoRR}, abs/1601.07539, 2016.

\bibitem{tatp}
A.~Wolski.
\newblock Tatp benchmark description (version 1.0), 2009.

\bibitem{Wu13}
E.~Wu and S.~Madden.
\newblock Scorpion: Explaining away outliers in aggregate queries.
\newblock In {\em PVLDB}, 2013.

\bibitem{Yates10}
J.~Yates.
\newblock Data entry error wipes out life insurance coverage.
\newblock In {\em Chicago Tribune}, 2005.

\end{thebibliography}
}

\techreport{\newpage
\appendix

\section{A Learning-based Approach}
\label{sec:heuristic}
  
A drawback of the MILP approach is that the generated models grow with the 
size of the database and query log.
However, we argue that the encoded information is necessary in order to generate a sufficient set of constraints that result in a good repair.
In this section, we examine an alternative, simpler, decision tree-based approach called \dt. 
We show that even in a simple case of a single query log and a complete complaint set, it is expected to perform poorly.
We will first describe how to model the repair process using a decision tree,
and then we will present and discuss experimental results that illustrate its limitations.

\subsection{Modeling Repairs with Decision Trees}

Rule-based learners are used in classification tasks to generate a set of rules, or conjunctive predicates that best classify a group of labeled tuples.
The rules are non-overlapping, and each is associated with a label---a tuple that matches a given rule is assigned the corresponding label.
These rules exhibit a natural parallel with SQL \texttt{WHERE} clauses, 
which can be viewed as labeling selected tuples with a positive label and rejected tuples with a negative label.
Similarly, the structure of the rules is identical to those that \sys is designed to repair.
Thus, given the database tuples labeled to describe the errors, we may use a rule-based learner to
generate the most appropriate \texttt{WHERE} clause.
We focus our attention on rule-based learners;
specifically, we experiment with the C4.5~\cite{quinlan1987} decision tree learner, which is an 
exemplar of rule-based learners.

A core limitation of this classification-based approach is that there is no means to 
repair \texttt{SET} clauses, which modify data values rather than simply label them.
We resolve this with a two step approach.
We first use the decision tree to generate a repair for the
\texttt{WHERE} clause, and then use the modified query to identify repairs for the \texttt{SET} clause.
The need for this two step procedure limits this approach to encoding and repairing at most one query
at a time.

\noindent
\textbf{Repairing the WHERE Clause:}
The \texttt{WHERE} clause of an update query is equivalent to a
rule-based binary classifier that splits tuples into two groups:
(1)~tuples that satisfy the conditions in the \texttt{WHERE} clause
and (2)~tuples that do not. A mistake in a query predicate can 
cause a subset of the tuples to be misclassified, and in turn,
translate into data errors. 
Therefore, repairing the complaints corresponds to repairing the imprecise classification. 

The repair works as follows: For an incorrect query $q$, let
$D_0$ be the database state before $q$, and $D_1^*$ the \emph{correct}
database state that should have been the result after $q$, if $q$ were correct.
We use each tuple $t \in D_0$ as an element in the input training data
for the classifier where the values (of each attribute) of $t$ define
the feature vector and the label for $t$:
	\[
    label(t)= 
    \begin{cases}
    true & \textrm{if\ }D_0.t \neq D_1^*.t\\
    false              & \text{otherwise}
    \end{cases}
\]
The \texttt{true} rules generated by the decision tree trained on this labeled dataset 
forms a disjunction of rules that constitute the repaired \texttt{WHERE} clause.

\noindent
\textbf{Repairing the SET Clause:}
The \texttt{WHERE} clause repair proposed by the classifier may not completely repair 
the complaints if there was also an error in the \texttt{SET} clause. 
In this case, we execute a second repair step.
\begin{figure}[!htb]
\centering
  \begin{subfigure} [t]{.75\columnwidth}
  \includegraphics[width = \columnwidth]{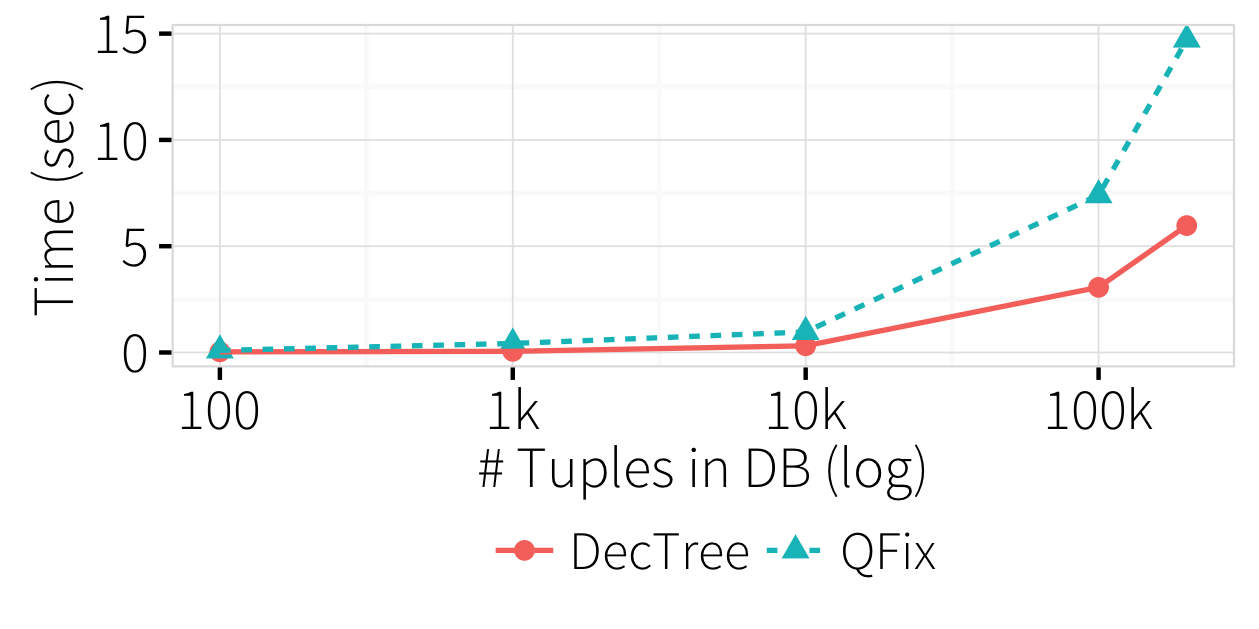}
  \caption{Comparison on Performance.}
  \label{f:heuristic_time} 
  \end{subfigure}\\

  \begin{subfigure} [t]{.75\columnwidth}
  \includegraphics[width = \columnwidth]{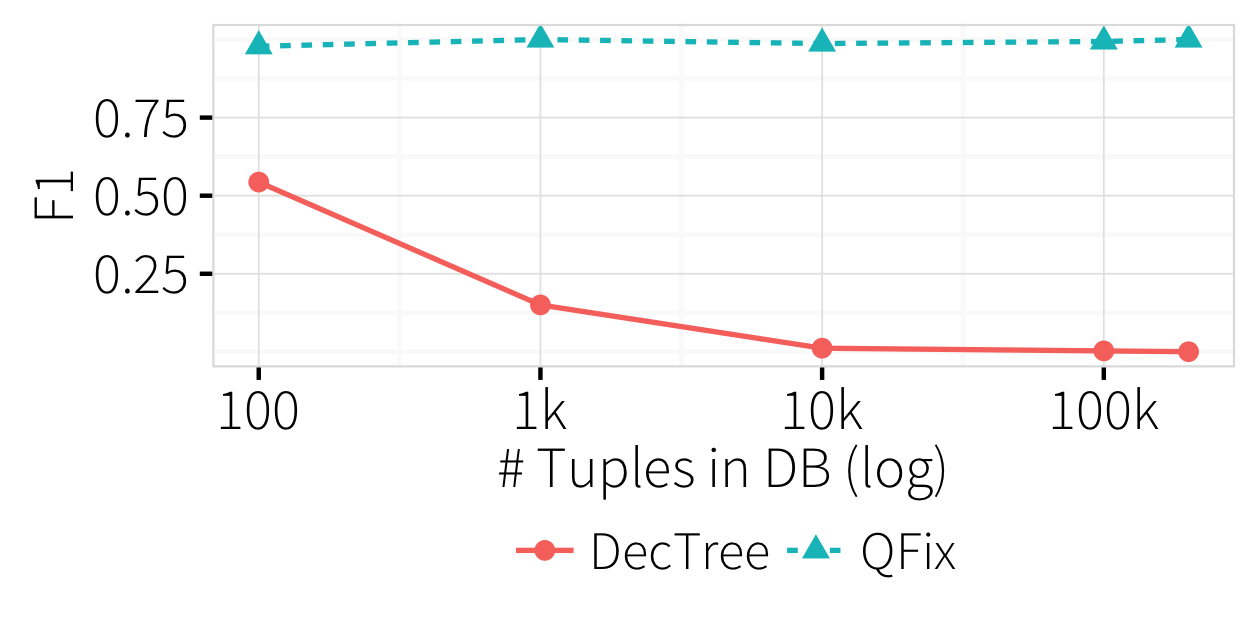}
  \caption{Comparison on Accuracy.}
  \label{f:heuristic_acc} 
  \end{subfigure}
 \caption{\dt compared with \sys}
 \label{f:heuristic}
\end{figure}

We model the errors as a simple linear system of equations: 
each expression in the \texttt{SET} clause is translated into a
linear equation in the same fashion as described in Section~\ref{sec:sol}.
Directly solving the system of equations for the undetermined variables 
will generate the desired repair for the \texttt{SET} expression.

\subsection{Experimental Results}

To illustrate these shortcomings, we compare \dt with \sys using a simplified version of the setup from Section~\ref{sec:experiments} that favors \dt.
We restrict the query log to contain a single query that is corrupted, use a complete complaint set  and vary the database size.
We use the following query template, where all \texttt{SET} clauses assign the attributes to constants,
and the \texttt{WHERE} clauses consist of range predicates:

{\scriptsize
\begin{verbatim}
  UPDATE table
  SET  (a_i=?), ...
  WHERE a_j in [?,?+r] AND ...
\end{verbatim}
}

Figure~\ref{f:heuristic_time} shows that although the runtime performance of \dt is better than \sys by small a constant factor ($\sim 2.5 \times$),
both runtimes degrade exponentially.
In addition, the \dt repairs are effectively unusable as their accuracy is low: the F1-score starts at $0.5$ and rapidly degrades towards $0$.
From these empirical results, we find that \dt generates low-quality repairs even under the simplest conditions---an approach
that applies \dt over more queries is expected to have little hope of succeeding.

There are three important reasons why \dt, and any approach that focuses on a single query at a 
time\footnote{Although our incremental approach tries to generate a repair for a single
query at a time, it encodes all subsequent queries in the log.}, will not perform well.

\begin{itemize}[itemsep=1pt, leftmargin=5mm]
    
\item \textbf{Single Query Limitation: }
In principle, one could attempt to apply this technique to the
entire log one query at a time, starting from the most recent query.
Even ignoring the low repair accuracy shown in Figure~\ref{f:heuristic_acc},
this approach is infeasible.
Consider that we generate a labeled training dataset to repair $q_i$ 
using the query's input and output database states $D_{i-1}$ and $D_i^*$.
Note that $D_i^*$ is the theoretically \emph{correct} database state assuming no errors in the query log.
We would need to derive $D_i^*$ by applying the complaint set to $D_n$ to create $D_n^*$, and roll back the database state.
Unfortunately, \texttt{UPDATE} queries are commonly surjective such
that their inverses are ambiguous, which means that it is often
impossible to derive $D_i^*$. In contrast, the incremental version of
\sys can bypass this problem by encoding subsequent queries in the log
in a MILP representation.

\item \textbf{Structurally Different \texttt{WHERE} Clause Results: } 
The basic classifier approach simply learns a set of rules to minimize
classification error, and can derive a clause whose struture is arbitrarily 
different from the original query's \texttt{WHERE} clause.
Although it may be possible to incorporate a distance measure as part of the decision tree
splitting criteria, it is likely to be a heuristic with no guarantees.

\item \textbf{High Selectivity, Low Precision: }
Classifiers try to avoid overfitting by balancing the complexity of the rules with classification accuracy.
This is problematic for highly selective queries (e.g., primary key updates), because the classifier
may simply ignore the single incorrect record and generate a rule such as \texttt{FALSE}.
In fact, this form of severely imbalanced data continues to be a challenge for most major classification algorithms~\cite{he2009learning, galar2012review}. 
Thus, we believe that alternative classification algorithms would not improve on these results. 
Compound with the fact that many workloads are primarily composed of
key update queries~\cite{oltpbench} this issue severely limits the
applicability of learning-based approaches.

\end{itemize}}

\end{document}